\documentclass[twocolumn]{aastex631}

\usepackage{amsmath}

\begin{document}
%TC:ignore
%\linenumbers
\title{Velocity-space turbulent cascade in the near-Sun solar wind: first insights from the Parker Solar Probe mission}

\author[0000-0002-7653-9147]{A. Larosa}
\affiliation{Istituto per la Scienza e Tecnologia dei Plasmi (ISTP), Consiglio Nazionale delle Ricerche, I-70126 Bari, Italy}

\author[0000-0002-7638-1706]{O. Pezzi}
\affiliation{Istituto per la Scienza e Tecnologia dei Plasmi (ISTP), Consiglio Nazionale delle Ricerche, I-70126 Bari, Italy}

\author[0000-0002-4625-3332]{T. Bowen}
\affiliation{University of California, Berkeley, CA, USA}
\affiliation{Space Sciences Laboratory, University of California, Berkeley, CA 94720-7450, USA}

\author[0000-0002-5981-7758]{L. Sorriso-Valvo}
\affiliation{Istituto per la Scienza e Tecnologia dei Plasmi (ISTP), Consiglio Nazionale delle Ricerche, I-70126 Bari, Italy}
\affiliation{Space and Plasma Physics, School of Electrical Engineering and Computer Science, KTH Royal Institute of Technology, 11428, Stockholm, Sweden}

\author[0000-0002-1128-9685]{N. Sioulas}
\affiliation{University of California, Berkeley, CA, USA}
\affiliation{Space Sciences Laboratory, University of California, Berkeley, CA 94720-7450, USA}

\author[0000-0002-5272-5404]{F. Pucci}
\affiliation{Istituto per la Scienza e Tecnologia dei Plasmi (ISTP), Consiglio Nazionale delle Ricerche, I-70126 Bari, Italy}

\author[0000-0002-0608-8897]{D. Trotta}
\affiliation{European Space Agency (ESA), European Space Astronomy Centre (ESAC), Camino Bajo del Castillo s/n, 28692 Villanueva de la Cañada, Madrid, Spain}
\author[0000-0003-1138-652X]{J. L. Verniero}
\affiliation{Heliophysics Science Division, NASA Goddard Space Flight Center, Greenbelt, MD 20771, USA}

\author[0000-0002-0396-0547]{R. Livi}
\affiliation{University of California, Berkeley, CA, USA}
\affiliation{Space Sciences Laboratory, University of California, Berkeley, CA 94720-7450, USA}

\author[0000-0003-0896-7972]{S. Bharati Das}
\affiliation{Center for Astrophysics | Harvard \& Smithsonian, 60 Garden Street, Cambridge, MA 02138, USA.}

\author[0000-0001-8478-5797]{A. Chasapis}
\affiliation{Laboratory of Atmospheric and Space Physics, University of Colorado, Boulder, CO 80303, USA}

\author[0000-0003-1059-4853]{D. Perrone}
\affiliation{ASI – Italian Space Agency, via del Politecnico snc, I-00133 Rome, Italy}

\author[0000-0002-1296-1971]{F. Valentini}
\affiliation{Dipartimento di Fisica, Universita` della Calabria, I-87036 Rende (CS), Italy}

\author[0000-0001-8184-2151]{S. Servidio}
\affiliation{Dipartimento di Fisica, Universita` della Calabria, I-87036 Rende (CS), Italy}
%TC:endignore

\begin{abstract}
In space plasmas, the rarity of collisions leads to complex structures in the velocity space where a turbulent cascade of the velocity distribution function fluctuations is thought to occur. Previous studies have explored this phenomenon using the Hermite decomposition of the ion velocity distribution function (VDF) in both magnetosheath data and numerical simulations. 
In this work, we investigate the Hermite spectrum of the ion VDFs measured by Parker Solar Probe in the inner heliosphere. We analyze a superalfvénic stream at a radial distances of $R \approx 28 R_{sun}$ and a subalfvénic at  $R \approx 11 R_{sun}$, the former characterized by a prevalence of VDFs with suprathermal beams (also known as hammerhead). The Hermite analysis is also compared with various proxies of energization and dissipation, in order to establish a connection between turbulent cascades in real space and those in the velocity space. A qualitative agreement between the energization proxies and the Hermite analysis is observed. The results are suggestive of the presence of a dual cascade in real and velocity space.  
\end{abstract}

\keywords{ heliosphere --- solar wind --- turbulence}

\section{Introduction} \label{sec:intro}
Nearly-collisionless turbulent space plasmas, such as the solar wind and Earth's magnetosheath, are usually far from the local thermodynamical equilibrium (LTE) \citep{bruno2013LRSP...10....2B, cassak2023}. Energy conversion and dissipation influence the entire six-dimensional phase space, and plasma velocity distribution functions (VDFs) are typically characterized by distinct non-Maxwellian structures, including temperature anisotropy, beams, rings, and finer-scale non-thermal velocity-space distortions.
The generation of finer and finer scale non-equilibrium structures in the VDF has often been envisioned as a turbulent cascade process in velocity space, where the cascading quantities are enstrophy and entropy \citep{Shecko2008PPCF...50l4024S, Tatsuno2009PhRvL.103a5003T, Shecko2016JPlPh..82b9012S, Servidio2017PhRvL.119t5101S}. Such a cascade is considered to be an avenue towards irreversible heating in plasmas with an extremely small collisional frequency.

Recently, numerical works \citep{Cerri2018ApJ, Pezzi2018PhPl...25f0704P, Pezzi2019ApJ, Cerri2021ApJ, Celebre2023PhPl, Nastac2024PhRvE.109f5210N} and observation studies in the near-Earth solar wind and in Earth's magnetosheath \citep{Servidio2017PhRvL.119t5101S, Wu2023ApJ...951...98W} have further explored this concept, while it still remains relatively unexplored in laboratory plasmas \citep{HowesLab2018PhPl...25e5501H} and in the inner heliosphere, mainly due to the lack of data. However, thanks to the NASA spacecraft Parker Solar Probe (PSP) \citep{FoxVelli2016, Nour2023SSRv..219....8R}, it has been possible to have a privileged observational point of a completely unexplored environment. PSP data have revealed, in the near-Sun solar wind, unprecedented complexity of the VDFs, which often exhibit the so-called ``hammerhead'' shape \citep{Verniero2020ApJS, Verniero2022ApJ}. 

A meaningful way to quantify this complexity is through the Hermite decompositions of the VDFs. The Hermite transform, which is a series expansion of the VDF in term of a Maxwellian weighted by Hermite polynomials, relates to a Maxwellian just as the Fourier transform relates to a plane wave: a single Fourier mode describes a plane wave, and likewise a single Hermite mode describes a Maxwellian. The more Hermite modes are non-zero the more the VDF is far from a Maxwellian.

This method has been used to investigate the phase space cascade in the magnetosheath \citep{Servidio2017PhRvL.119t5101S}, in solar wind reconnection exhausts \citep{Wu2023ApJ...951...98W}, and in Vlasov-Maxwell numerical simulations exploring different turbulent fluctuations amplitude and plasma $\beta$ conditions \citep{Pezzi2018PhPl...25f0704P, Cerri2018ApJ}. Different heating mechanism have different velocity space signatures \citep{HuangTTDvsLD2024JPlPh..90d5301H} whose dominance strongly depends on the plasma, kinetic to magnetic pressure ratio, $\beta$ \citep{Cerri2021ApJ, HuangTTDvsLD2024JPlPh..90d5301H}. 
Therefore, it is crucial to understand if and how the velocity-space turbulent cascade also depends on $\beta$. 

The Hermite or Hermite-Laguerre transform also provides a nonparametric and analytical representation of the VDFs, which enables one to compute VDF derivatives, necessary to study wave-particle interactions \citep{Bowen2022PhRvL,Coburn2024ApJ}. 

Moreover, the exponent of the power-law Hermite spectrum might be informative about the dominant physical process that controls the cascade. According to a Kolmogorov-like phenomenology, a $-1.5$ exponent is expected for a phase-mixing or electric-field-dominated regime, while a $-2$ exponent indicates a magnetic-field-dominated regime \citep{Servidio2017PhRvL.119t5101S}. 
Furthermore, the exponent has been shown to be affected by the competition of linear against nonlinear processes, with a steeper spectrum in the nonlinear-dominated regime in electrostatic ion-temperature-gradient driven drift-kinetic turbulence \citep{Parker2016PhPl...23g0703P}, while in Vlasov-Poisson systems power-laws are observed when instabilities or turbulence are present, but not in linear regimes \citep{Celebre2023PhPl}.

The study of the phase space cascade through the Hermite spectrum can be complemented by inspecting different measures of energy conversion and dissipation, which can be based either on the energy in the fields or on the VDF. 
This is crucial in order to link the velocity-space complexity with the turbulent cascade. Indeed, the existence of a cascade in the phase space and its relation to turbulent fluctuations are crucial elements towards the understanding of how nearly collisionless plasmas are heated in a thermodynamic (irreversible) sense \citep{Shecko2008PPCF...50l4024S,Tatsuno2009PhRvL.103a5003T,2016PhRvL.116n5001P, Nastac2024PhRvE.109f5210N}. 

In this work, we explore this link through a study of the phase-space cascade using PSP measurements in the inner heliosphere. In particular, we focus on a proxy for local turbulent energy transfer ($LET$) \citep{Sorriso2019PhRvL.122c5102S} and on a quantitative measure of the deviation of the VDFs from Maxwellianity \citep{kp2009JGRA, Liang2020JPP}. Although these measures do not distinguish between different physical processes, they are capable of locating regions of strong dissipation and non-linear energy transfer that should also be captured in the Hermite analysis of the VDFs. An extensive review of these quantities is given in \cite{Pezzi2021MNRAS}.

Data and methods are described in Section~\ref{sec:data_methods}, the properties of the streams under investigation and their Hermite spectra in Section~\ref{sec:streams}, and the link between VDF and fields-based proxies for dissipation and energization in Section~\ref{sec:FieldVsVDF}. We conclude in Section~\ref{sec:Discussion}.

\section{Data and Methods} \label{sec:data_methods}
We use data from the FIELDS \citep{Bale2016} and SWEAP \citep{Kasper2016} instrument suites onboard PSP. FIELDS provides the magnetic field data from both the flux-gate magnetometer and the search coil magnetometer \citep{Jannet2021JGRA..12628543J, Thierry2022JGRA..12730018D}. These two data products are combined in the SCaM dataset \citep{Bowen2020SCamJGRA..12527813B} that has an excellent signal-to-noise ratio. Unfortunately, due to an anomaly, only two components of the search coil magnetometer magnetic field are available past the first encounter \citep{Thierry2022JGRA..12730018D}.

The SWEAP sensors provide the electron pitch-angle distribution (ePAD) from the SPAN-e instrument \citep{phyllisSpanE2020} and the proton velocity distribution functions and their moments from the SPAN-i instrument \citep{Livi2022ApJ}. 
The ePAD is used to select intervals with a unidirectional electron strahl in order to exclude complex magnetic topologies from the analysis \citep[see][for the connections between the strahl and the magnetic topology]{OwensFors2013LRSP...10....5O}, that could invalidate the interpretation of the time averages as ensemble averages.

In this work we use SPAN-i data from encounters 4 and 18, whose respective cadence is 7 and 1.7 seconds corresponding to about 13 and 21 cyclotron times, being the ion cyclotron frequency $\Omega_{cp}=e B/m_pc$ computed with the average magnetic field for each interval. SPAN-i VDFs have a three-dimensional (3D) resolution in velocity space of $8 \times32 \times8$ in azimuth, energy and elevation, respectively.   
For each bin of the SPAN-i instrument grid, we move from the instrument grid to the field-aligned coordinates ($v_{\parallel}$, $v_{\perp 1}$, $ v_{\perp 2}$), where $v_{\parallel}$ is along the local magnetic field and $v_{\perp 1}$ and $ v_{\perp 2}$ are orthogonal to it. The local magnetic field used is the one at the SPAN-i cadence available in the L3 SPAN-i data products.

Then, we define $v_{\perp}= \sqrt{v_{\perp 1}^2 + v_{\perp 2}^2}$, thus assuming gyrotropy \citep[as in ][]{Bowen2022PhRvL} in the directions perpendicular to the local field. Hence, we reduce the VDF to a two-dimensional (2D)  velocity-space in $v_{\parallel}$ and $v_{\perp}$.

The Hermite decomposition requires defining the velocity domain in $(-\infty,\infty)$. Therefore, following \citet{Bowen2022PhRvL}, we impose $f(-v_{\perp}) = f(v_{\perp})$ and extend the grid to negative values in the perpendicular direction consistently with the gyrotropy assumption (see Appendix~\ref{app:gyro} to visualize the effect of this approach on the VDFs). Note that such a procedure implies null odd Hermite coefficients in the perpendicular direction.

We shift the velocity grids, in both parallel and perpendicular directions, such that the bulk speed is zero. Moreover, the velocity grids are normalized by their respective (perpendicular and parallel) thermal speeds. The shift and the normalization ensure that fluctuations in the $1^{\rm st}$ and $2^{\rm nd}$ order VDF moments, associated with bulk flows or large-scale energy conversion (e.g., due to radial expansion) do not influence the Hermite spectrum at high modes \citep{Pezzi2018PhPl...25f0704P}.

\begin{figure*}[!htb]
    \centering
    \includegraphics[width=0.7\textwidth]{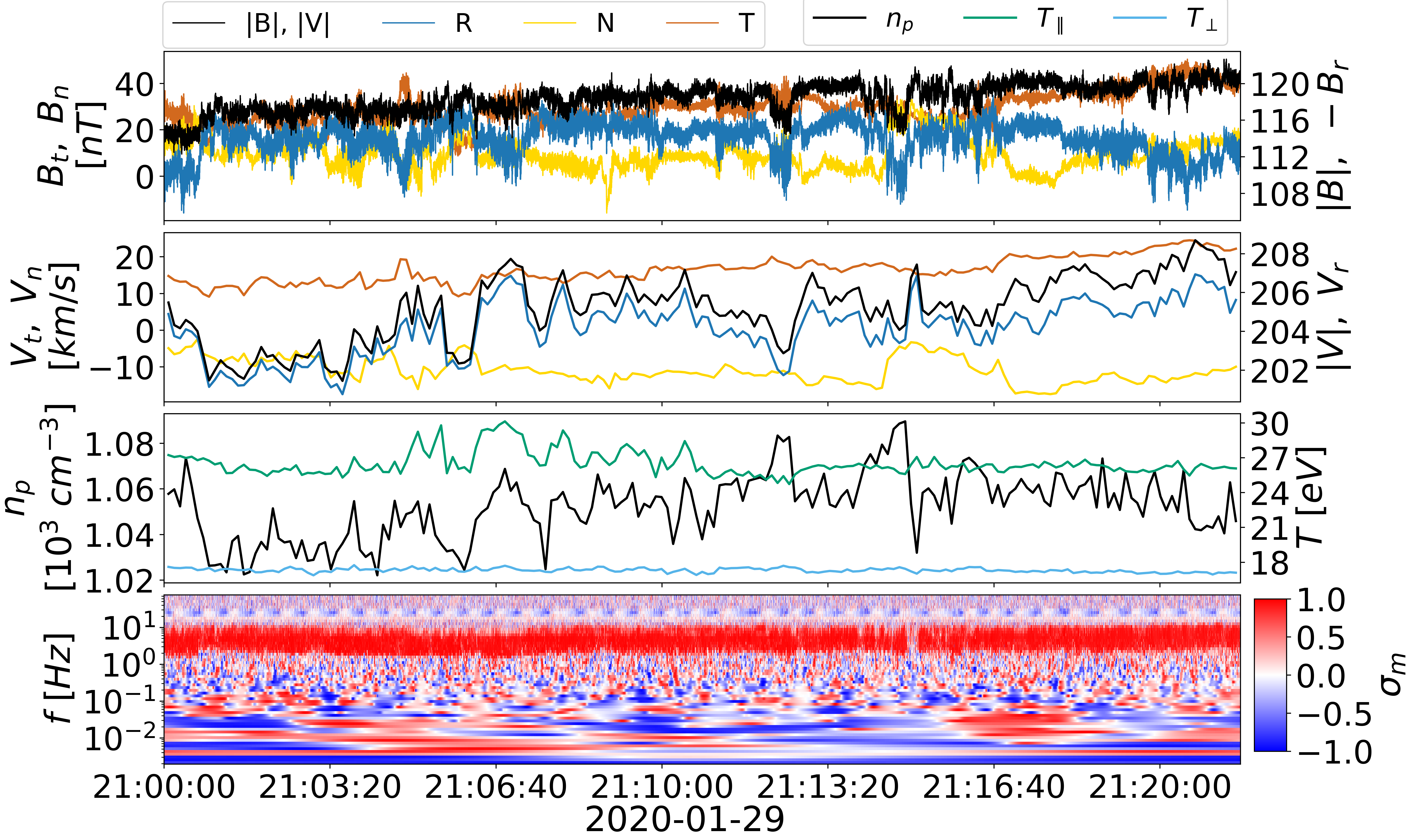}
    \includegraphics[width=0.7\textwidth]{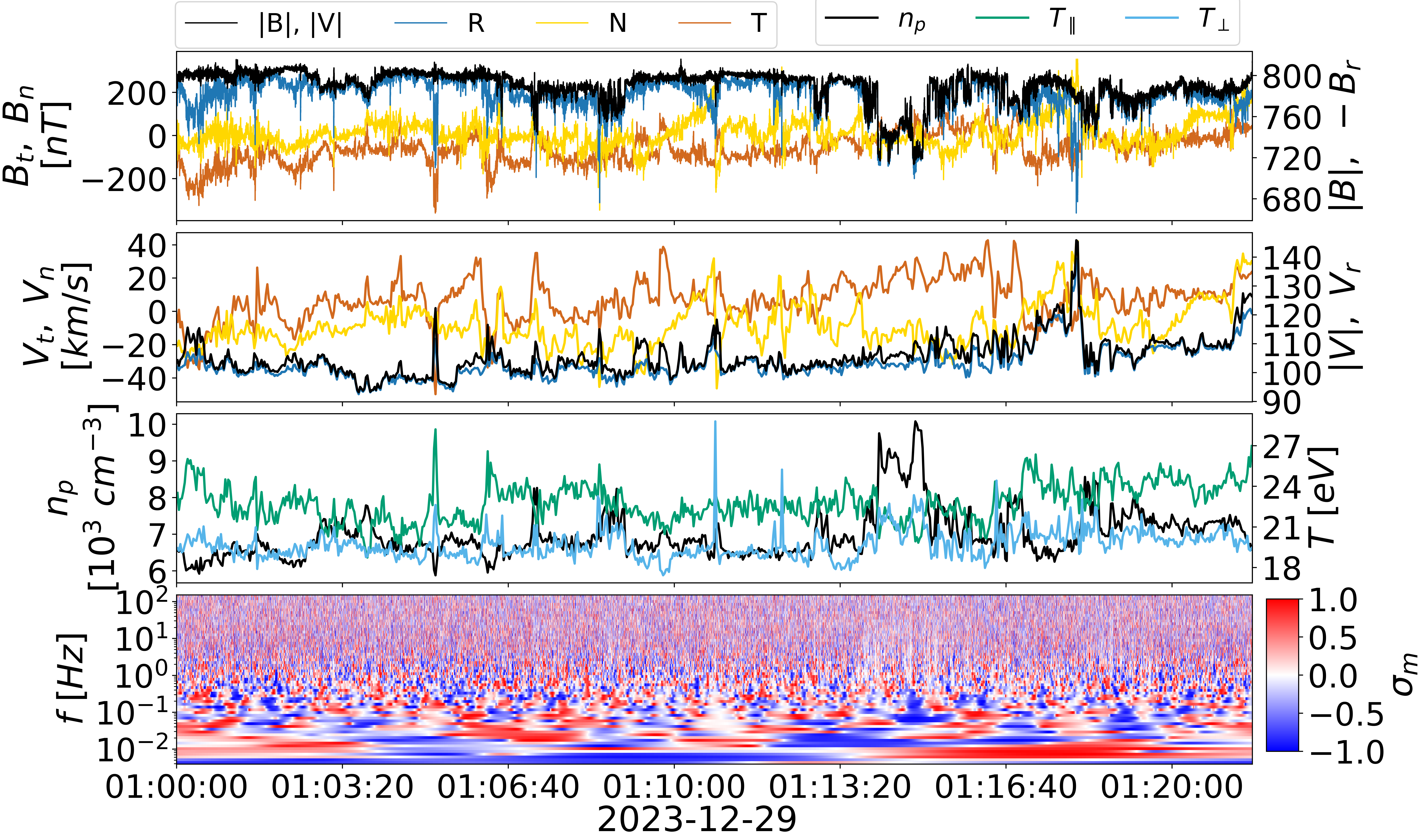}
    \caption{Magnetic field, velocity field, density and parallel and perpendicular temperatures and magnetic helicity in the \textit{wave stream} (top) and \textit{turbulent stream} (bottom).}
    \label{fig:stream}
\end{figure*}

The Hermite decomposition of a VDF, at a given instant in time, is hence defined as:
\begin{equation} \label{eq:decomp}
    f(v_\perp, v_\parallel) = \sum_{m_\perp, m_\parallel} g_{m_\perp, m_\parallel} \, \psi_{m_\perp}(\xi_\perp) \, \psi_{m_\parallel}(\xi_\parallel),
\end{equation}
where the normalized velocity coordinates are $\xi_\perp = (v_\perp - u_\perp)/v_{\text{th},\perp}$ and $\xi_\parallel = (v_\parallel - u_\parallel)/v_{\text{th},\parallel}$, being $u$ and $v_{\text{th}}$ respectively the bulk velocity and the thermal speed in each direction. The orthonormal Hermite  eigenfunctions are
\begin{equation}
    \psi_m(\xi) = \frac{H_m(\xi)}{\sqrt{2^m \sqrt{\pi} m!}} e^{-\frac{\xi^2}{2}},
\end{equation}
where $H_m(\xi)$ are the physicist’s Hermite polynomials
\begin{equation}
    H_m(\xi) = (-1)^m e^{\xi^2} \frac{d^m}{d\xi^m} e^{-\xi^2},
\end{equation}
and $m=m_\perp, m_\parallel$. By exploiting the orthonormality condition $\int_{-\infty}^{\infty} d\xi \psi_m(\xi) \psi_n(\xi) = \delta_{m,n}$, the Hermite coefficient $g_{m_\perp, m_\parallel}$ can be calculated as:
\begin{equation} \label{eq:gmn}
    g_{m_\perp, m_\parallel} = \int_{-\infty}^{\infty} \int_{-\infty}^{\infty} f(v_\perp, v_\parallel) \, \psi_{m_\perp}(\xi_\perp) \, \psi_{m_\parallel}(\xi_\parallel) \, d\xi_\perp \, d\xi_\parallel.
\end{equation}

As discussed by \citet{Servidio2017PhRvL.119t5101S} and \citet{Pezzi2018PhPl...25f0704P}, each  VDF is interpolated on the 2D quadrature grid corresponding to the roots of the $N_v+1$-th Hermite polynomials before computing the coefficients $g_{m_\perp, m_\parallel}$, being $N_v$ the maximum order of Hermite polynomials in each direction. This procedure allows us to exploit the Gauss-Hermite quadrature to compute integrals and, in particular, the Parseval-Plancerel spectral theorem, $\int f^2 d\xi_\perp \, d\xi_\parallel = \sum g^2_{m_\perp, m_\parallel}$, which connects the enstrophy with the sum of Hermite coefficients squared. The one-dimensional Hermite spectrum $P(m)$ is then computed over concentric-shells of unit thickness in the Hermite space, i. e. $P(m) = \sum_{m-1/2< |{\bf m'}| \leq m+1/2} g_{{\bf m'}}^2 $ where $m=\sqrt{m_\perp^2+m_\parallel^2}$ and ${\bf m}= (m_\perp, m_\parallel)$. For this work, we choose $N_v=50$ such that the number of Hermite grid points ($50^2=2500$) is close to the number of instrument grid points ($8\times 32\times 8=2048)$.
We note that in general shell averaging the spectra could cause the mixture of high and low $m$ modes if the parallel and perpendicular thermal speed are very different, therefore using the trace speed should also be considered. However, we verified that for the streams studied here using the trace speed does not modify significantly the spectra.

\section{Streams properties and Hermite spectra} \label{sec:streams}
We investigate the properties of two different intervals, hereafter identified as the \textit{wave stream} (Figure~\ref{fig:stream}, top) and the \textit{turbulent stream} (Figure~\ref{fig:stream}, bottom). Figure~\ref{fig:stream} shows, from top to bottom, the magnetic field components, the velocity components, proton density and parallel and perpendicular temperatures and the spectrogram of the magnetic helicity for the two streams. 
The magnetic helicity is
defined as 

\begin{equation}
\sigma_m = \frac{2 \Im \left(\tilde{B_T^{\star}}\tilde{B_N}\right)}{|\tilde{B_R}|^2 + |\tilde{B_T}|^2 + 
|\tilde{B_N}|^2},
\end{equation}

where $\mathrm B$ indicates the magnetic field components, the $\tilde{}$ represents the wavelet-
transformed quantities and $\star$ represents the operation complex conjugation~\citep{Matthaeus1982}. We use a Morlet wavelet.

The streams are chosen in order to have a mostly radial magnetic field in order to avoid field of view issue with SPAN-i \citep[see][]{Livi2022ApJ}.
The average properties are listed in Table~\ref{tab:summary}. The superalfvénic \textit{wave stream} was measured on 2020 January 29 from 21:00 to 21:21:37 UTC and the subalfvénic \textit{turbulent stream} on 2023 December 29 from 01:00 to 01:21:37 UTC.

\begin{table}[!htb]
    \centering
    \begin{tabular}{|c|c|c|}
        \hline
        \textbf{Quantity} & \textit{wave} & \textit{turbulent} \\
        \hline
        $B_{\rm rms}~[nT]$ & 9.83 & 95.81 \\
        $B_{\rm rms}/\langle B\rangle$ & 0.08 & 0.12 \\
        $\langle \beta_{p}\rangle$ & 0.6 & 0.1 \\
        $\langle \theta_{BV}\rangle^\circ$ & 168 & 116 \\
        $\langle \sigma_{c}\rangle$ & 0.69 & 0.79 \\
        $R \ [R_{sun}]$ & 28.0 & 11.4 \\
        $M_A$ & 2.6 & 0.5 \\
        $f_c \ [Hz]$ & 1.8 & 12 \\
        \hline
    \end{tabular}
    \caption{Summary of streams average properties.}
    \label{tab:summary}
\end{table}

Both intervals are characterized by a mostly radial magnetic field and are highly Alfvénic and unbalanced, with high normalized cross-helicity $\sigma_c$, see Table~\ref{tab:summary}.
The rms fluctuation level and the density are one order of magnitude stronger for the \textit{turbulent stream} closer to the Sun, but the ratio $B_{\rm rms}/\langle B\rangle$ is of the same order for both intervals. $\langle B\rangle$ is the magnitude of the mean of the magnetic field over the full interval.
The average proton kinetic to magnetic pressure ratio $\beta_p = \frac{nk_BT_p}{B^2/2\mu_0}$ is smaller for the \textit{turbulent stream}. 

The two streams have a different average angle between the magnetic field and the velocity, $\theta_{BV}$. The \textit{wave stream} has almost antiparallel sampling ($\theta_{BV}=168^\circ$), while the \textit{turbulent stream} has more perpendicular sampling ($\theta_{BV}=116^\circ$). The angle is computed in the spacecraft frame to take into account the large tangential velocity of PSP which affect the sampling direction \citep[e.g.][]{Klein_2015ApJ...801L..18K}. This explains why, even though both the plasma velocity and the magnetic field are mostly along the radial direction (see Fig.\ref{fig:stream}) we can have large $\theta_{BV}$.

\begin{figure}
\includegraphics[width=0.45\textwidth]{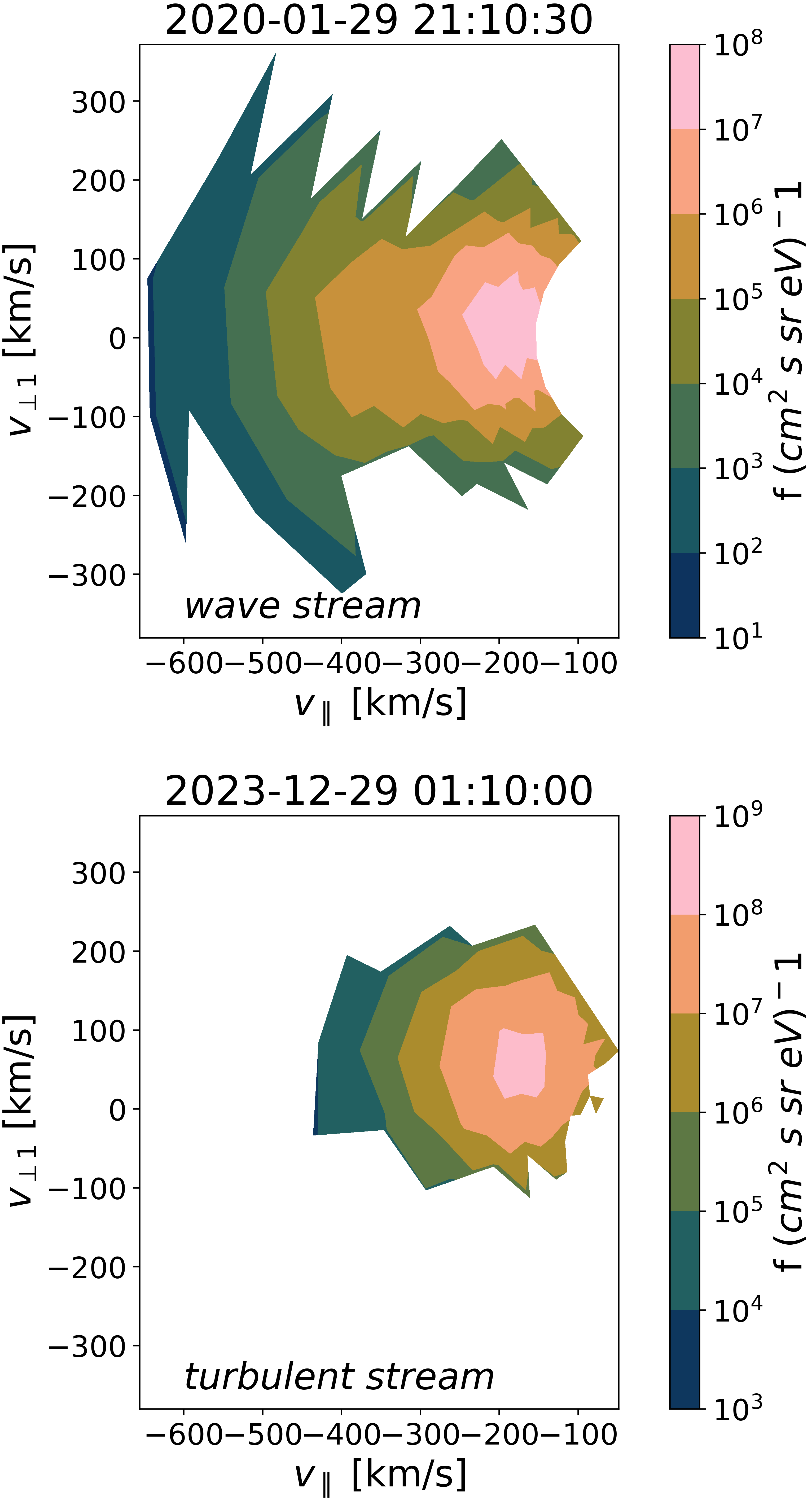}
    \caption{Representative VDFs for the \textit{wave stream} (top) and \textit{turbulent stream} (bottom). The VDFs are integrated along $\phi$ and plotted in the Energy-$\theta$ plane \citep[see][]{Verniero2020ApJS} in field aligned coordinates. These are the original non-gyrotropic PSP VDFs for which the procedure described in Sec~\ref{sec:data_methods} has not been applied yet.}
    \label{fig:vdfs}
\end{figure}

Two representative VDFs for the two streams are shown in Figure~\ref{fig:vdfs}. For both streams the SPAN-I sensor has a good field of view, with the core of the distribution being always mostly resolved. PSP observations have shown the prevalence of VDFs with large proton beams that undergo perpendicular velocity-space diffusion at higher energies, resembling a "hammerhead" shape \citep{Verniero2020ApJS, Verniero2022ApJ} (hereby referred to as hammerhead distributions). This is exemplified in Figure~\ref{fig:vdfs} (top), where the beam at larger $v_\parallel$ spreads to larger $v_\perp$. 
Hammerhead distributions are dominant in the \textit{wave stream}, which present parallel anisotropy $T_\parallel> T_\perp$ (as highlighted in the third panel of Figure~\ref{fig:stream}, top) and intense wave activity at about the ion-cyclotron frequency, as revealed by the magnetic helicity \citep{Pecora2021A&A...650A..20P, Bowen2022PhRvL, TrottaLarosa2024ApJ} in the bottom panel of the top window in Figure~\ref{fig:stream}. The resonant interaction between these waves, belonging to the fast magnetosonic/whistler branch, and the proton beam is likely the cause of the hammerhead distributions \citep{Verniero2022ApJ}.

In the \textit{turbulent stream} no hammerhead distributions are present (Figure~\ref{fig:vdfs}, bottom) and the temperature anisotropy is less pronounced. No distinct wave activity is observed in this stream. However, the lack of wave activity could also be due either to the wave polarization plane not being properly sampled or to stronger turbulent fluctuations when $\theta_{BV}$ is perpendicular \citep{Bowen2020ApJS}. 

\begin{figure}
\includegraphics[width=0.45\textwidth]{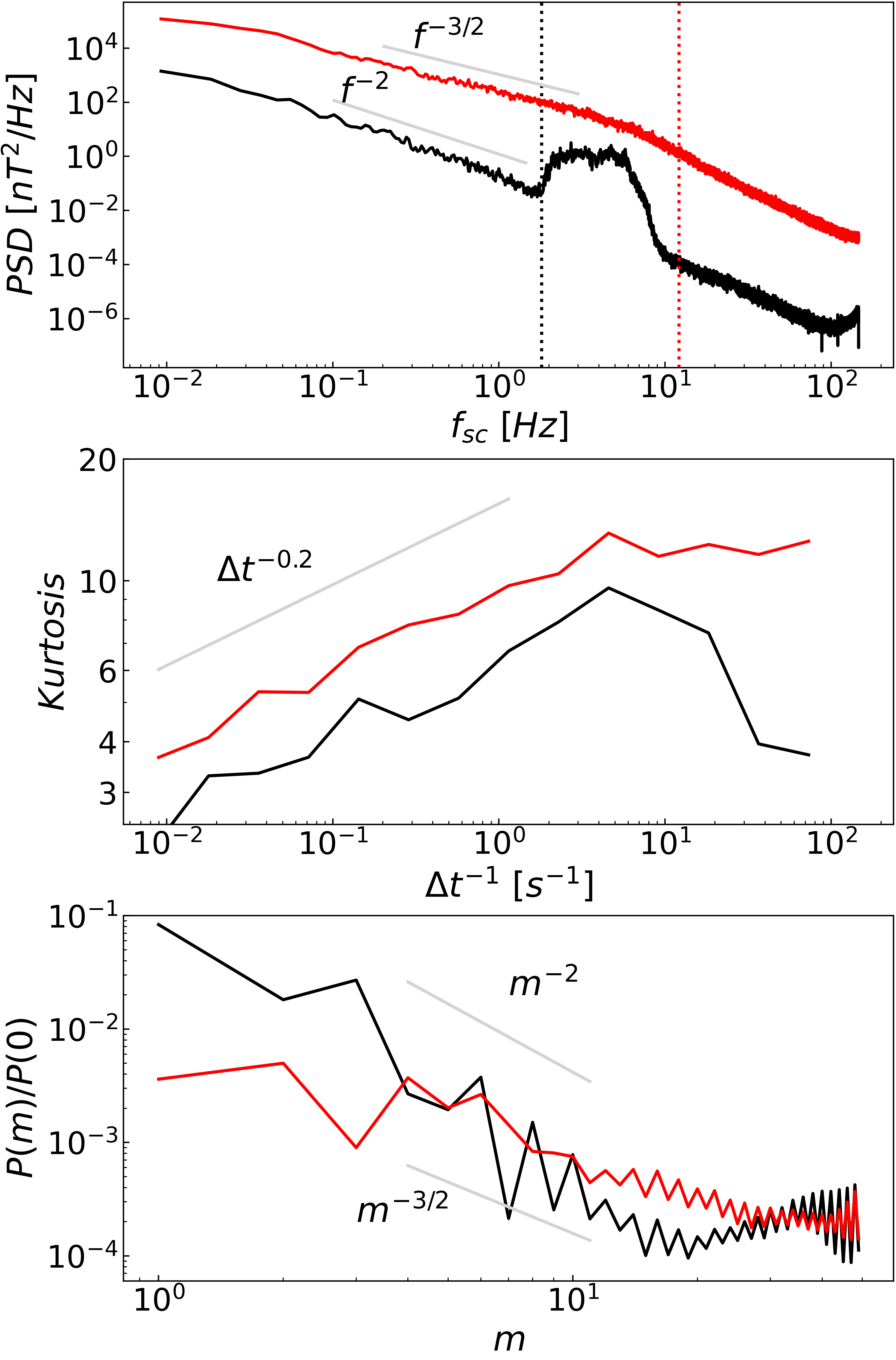}
    \caption{Top to bottom: Power spectral density and kurtosis of the magnetic field and average Hermite spectrum for the \textit{wave stream} (black) and \textit{turbulent stream} (red). In gray typical reference slopes (these are not fits) to aid the eye. The slopes for the Hermite spectra are the ones from the \cite{Servidio2017PhRvL.119t5101S} phenomenological model.}
    \label{fig:spectral_prop}
\end{figure}

The spectral properties of the two streams are highlighted in Figure~\ref{fig:spectral_prop}.
Since solar wind turbulence is anisotropic with respect to the ambient magnetic field, we expect the sampling direction to affect the fluctuations spectral properties \citep{Shebalin1983JPlPh..29..525S, Horbury2008PhRvL, podesta2009ApJ...698..986P, chen2010ApJ...711L..79C, Horbury2012SSRv, Chen2016JPlPh..82f5302C, Bowen2020ApJS, Sioulas2023ApJ...951..141S}. 
The top panel of Figure~\ref{fig:spectral_prop} shows the magnetic power spectral trace for the two intervals (black for the \textit{wave stream}, red for the \textit{turbulent stream}). The SCaM dataset is used for the former to avoid the flattening of the spectrum at high frequency present in the MAG data due to a low signal-to-noise ratio. The flattening is absent in the \textit{turbulent stream} for which we use the MAG data. Vertical dashed lines show the ion-cyclotron frequency for the two streams (same colors), while reference power laws are indicated in gray. 

The \textit{wave stream} shows a steeper exponent of the power-spectral density (PSD) of the magnetic field fluctuations, compatible with $-2$, compared to the \textit{turbulent stream} for which the spectral exponent is $-3/2$ \citep[e.g.,][]{Boldyrev2005ApJ...626L..37B, Boldyrev2006PhRvL..96k5002B}. To be more precise for the former the linear regression in the log-log space gives a slope of $-2.17 \pm 0.05$ in the range $[0.1, 0.9] \ Hz$  while for the latter $-1.50 \pm 0.01$ in the range $[0.1, 4] \ Hz$. 
The spectral slope difference could be attributed to different factors: i) the different sampling angle which makes sampling either quasi-perpendicular or quasi-parallel; ii) the different amplitude-to-ordered magnetic field which may induce a different type of turbulence (weak vs strong) in the two intervals \citep{2015JFM...770R...1M}, and iii) the spectral anisotropy which could steepen magnetic field spectra in the parallel direction \citep{2022JPlPh..88e1501S}. We also highlight that previous observations found evidence of a clear $-2$ spectrum at $1\,\mathrm{au}$  only if calculated with respect to the local mean field \citep{Chen2011MNRAS.415.3219C}. Here, we observe a similar slope, despite we adopt the global mean field, since $B_{\rm rms}/\langle B\rangle$ is small (Table~\ref{tab:summary}) and local- and global-mean field based analyses are expected to approach each other.

Notably, the \textit{wave stream} PSD presents a substantial bump around the ion-cyclotron frequency (dashed black line in the top panel of Fig.~\ref{fig:spectral_prop}) in agreement with the magnetic helicity signature in Fig.~\ref{fig:stream} (left).

The intermittency properties of the two intervals are evaluated using the scaling  of the kurtosis, $K(\Delta t)=\langle \Delta \mathbf{B}(t)^4\rangle/\langle \Delta \mathbf{B}(t)^2\rangle^2\sim \Delta t^{-\kappa}$, defined as the scale-dependent fourth-order moment of the distribution of the magnetic field fluctuations 
with $\Delta t$ defining a time scale, normalized to the squared second-order moment \citep{Frisch1995,Bruno2003}. 
The kurtosis is a measure of how fast the fluctuations distribution function tails go to zero, which is sensitive to how space filling the fluctuations at a given scale are.
An increase in $K$ toward smaller scales indicates higher distribution tails, so that the turbulent structures become confined to a progressively smaller fraction of space as the scale decreases.
Consequently, intermittency in a generic time series manifests as a monotonic (power-law) increase of the kurtosis with decreasing scale \citep{Frisch1995, Sioulas2022ApJ...934..143S,Mondal2025}.

In order to evaluate $\Delta \mathbf{B}$ we use the 5 points increments method

\begin{align*}
\Delta \mathbf{B}(t, \Delta t) 
= \frac{1}{\sqrt{35}} \big[ 
    \mathbf{B}(t - 2\Delta t) 
    - 4\mathbf{B}(t - \Delta t) 
    + 6\mathbf{B}(t) \\
    - 4\mathbf{B}(t + \Delta t) 
    + \mathbf{B}(t + 2\Delta t) 
\big]   
\end{align*}

which is more suitable for small scale increments since it can capture the proper scaling even when the spectrum is very steep (like in the transition range), contrary to the 2-points method \citep{Cerri2019FrASS...6...64C, Cho2019ApJ...874...75C, Nikos2024arXiv240404055S}.  

The kurtosis is first computed for each magnetic field component, and then averaged \citep{Sorriso-Valvo2023}, to provide an overall indication of intermittency. 
As illustrated in the center panel of Figure~\ref{fig:spectral_prop}, the kurtosis decreases as a power law of the time scale, confirming that a turbulent cascade is active in both streams \citep{Frisch1995}. The scaling exponent is roughly -0.2, comparable to typical values in the solar wind \citep{Sorriso-Valvo2023}, and indicating strong intermittency. However,  like for the magnetic power spectra, the kurtosis values are different in the two intervals. The larger $K$ observed at all scales for the \textit{turbulent stream} (red line) compared to the \textit{wave stream} (black line) are indicative of enhanced intermittent fluctuations (i.e., with more non-Gaussian distributions) . Additionally, while in the \textit{turbulent stream} the kurtosis flattens when approaching ion scales, the evident decrease in the \textit{wave stream} suggests that the observed waves decorrelate the turbulent strucutures, restoring the Gaussian statistic of the fluctuations and thus eliminating the intermittency. Indeed a negative correlation between the presence of ion cyclotron wave activity and the level of intermittency has been observed also in Solar Orbiter data \citep{CaroneF2021A&A...656A..16C}.

The Hermite decomposition, described in Section~\ref{sec:data_methods}, is applied to each VDF in the streams and the corresponding spectra, averaged over the full length of the intervals, are shown in the bottom panel of Figure~\ref{fig:spectral_prop}.
The oscillatory behavior in both spectra is due to the lack of power in the odd modes of the perpendicular spectrum caused by the gyrotropy assumption, which obviously affect also the 1D isotropic spectrum. The statistical error, computed as the standard deviation of the mean at each $m$, is too small to be distinguishable from the spectra. 

In the \textit{wave stream}, the Hermite coefficients for $m=1, 2, 3$ contain more power than in the turbulent case, due to the presence of the beam and hammerhead structure. Indeed, tests with synthetic data (not shown) illustrate that the presence of a dense beam at a few thermal speeds, and a much less dense beam at $\gtrsim 5$ thermal speeds provide power only to the first few Hermite modes. 
In contrast, the \textit{turbulent stream} velocity distribution functions exhibit a more Maxwellian core and velocity-space distortions at diverse velocity scales. The more Maxwellian core implies that the coefficient $m = 0$ has a higher power, which is part of the reason for the observed differences between the two streams at low $m$. Note that a similar behavior in the Hermite spectra for $1 \leq m \leq 4$ has been observed in Solar Orbiter data for the VDF of a reconnection exhaust with respect to the typical core-beam VDF of the ambient solar wind \citep{Wu2023ApJ...951...98W}.
At high $m$ ($8 \lesssim m \lesssim 15$), the Hermite spectrum shows more power in the \textit{turbulent stream} compared to the \textit{wave stream}. This difference can be interpreted as an effect of the stronger velocity-space cascade in the \textit{turbulent stream}, which induces fine-scale VDF distortions. 

The spectral slope computed in the range $4 \leq m < 12$, is about $-2$ for both streams.
The range of $m$ used to compute the Hermite spectral exponent is selected to ensure a sufficiently populated spherical shell while avoiding the high-$m$ regime dominated by the noise. 
The slopes are consistent with the expectations for a low-$\beta$ plasma \citep{Servidio2017PhRvL.119t5101S}, but this agreement should be considered only qualitative considering the limitation imposed by the resolution of the instrument and the limited spectral range over which they are computed. Indeed the slope is very sensitive to the chosen range of $m$.

We tested the spectra against the ones built on the one-count VDFs and the spectra of bi-Maxwellian VDFs interpolated on SPAN grid. In both cases, the spectra in Figure~\ref{fig:spectral_prop} were above these signals interpreted as noise levels (see Appendix~\ref{app:biMax}).
We purposefully did not discuss the spectra for $m \gtrsim 20$ because the flattening is likely due to the interpolation errors. This, however, does not affect Hermite modes at lower $m$.

\section{Transfer and conversion of energy} 

\label{sec:FieldVsVDF}
\begin{figure*}[htbp]
    \centering

\includegraphics[width=0.7\textwidth]{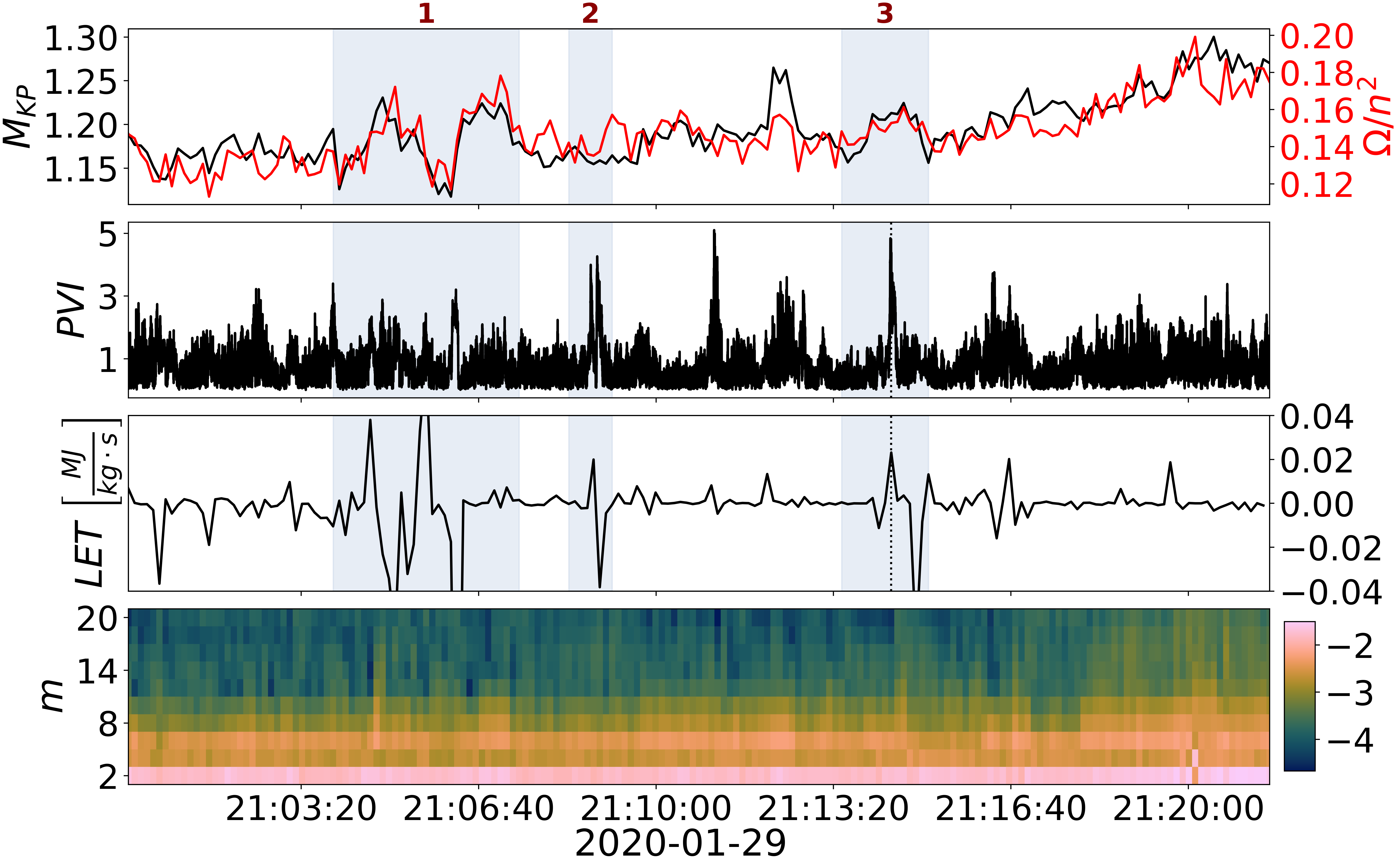}
\includegraphics[width=0.7\textwidth]{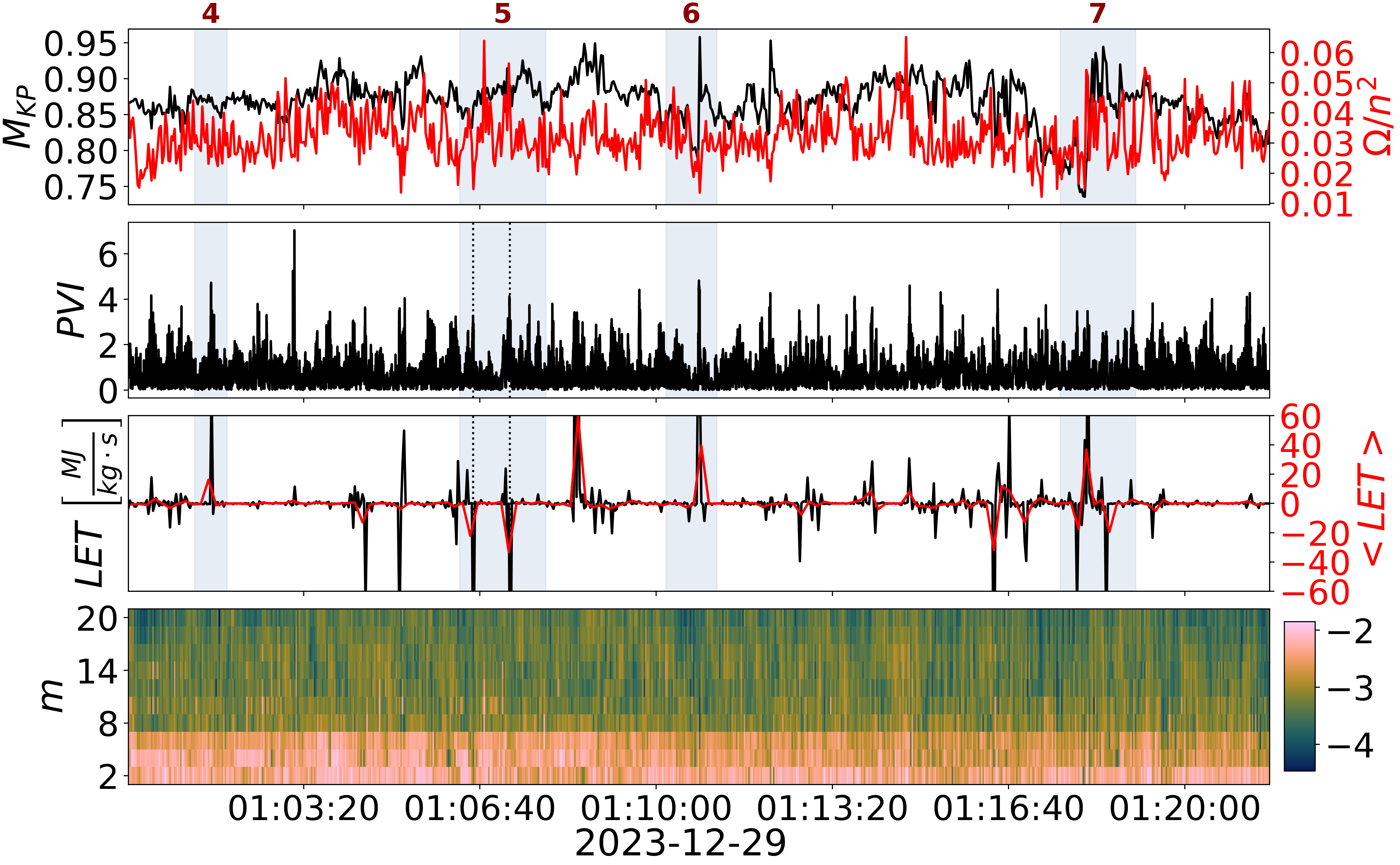}
    \label{fig:NoHornet}
    \caption{From top to bottom: Kaufman-Paterson measure and normalized enstrophy, $PVI$, $LET$, and Hermite spectrogram for the \textit{wave stream} (top) and the \textit{turbulent stream} (bottom). The shaded areas highlight regions of interest described in the main text. The Hermite spectrograms are represented through the scientifically derived color map batlow \citep{CrameriNatureComm2020}.}
\end{figure*}

In order to gain some insights on the ongoing energy transfer and dissipation, we compare the different field- and VDF-based diagnostics to highlight energy conversion \citep[see, e.g.,][and references therein]{Pezzi2021MNRAS}. 

VDF-based proxies estimate how far the VDF is from the equilibrium Maxwellian distribution. Here, we adopt the Kauffman-Paterson measure \citep{KP2009JGRA..114.0D04K, Liang2020JPP} 
\begin{equation}
    M_{KP} (t) = \frac{s_M(t)-s(t)}{(3/2)k_B n},
\end{equation}
where $s(t)=-k_B \int d^3v f(\textbf{v}, t) \ln f(\textbf{v}, t)$ and $n$ is the proton density, and the enstrophy \citep{Servidio2017PhRvL.119t5101S,Pezzi2018PhPl...25f0704P}
\begin{equation}
 \Omega (t) = \sum_{m>0} [g_m(t)]^2 .
\end{equation}
The enstrophy is related to the Maxwellianity indicator $\epsilon = \frac{1}{n} \sqrt{\int (f-f_{M})^2d^3v}$ through $\Omega = \epsilon^2 n^2$, where $f_{M}$ is the Maxwellian built using the local density and trace temperature \citep{Pezzi2018PhPl...25f0704P}.

The Kauffman-Paterson measure evaluate how far the entropy of the measured distribution $s(t)$ is from the entropy $s_M(t)$ of the local Maxwellian $f_{M}$. A value close to zero indicate that the plasma is close to local thermodynamical equilibrium (LTE) while, conversely, as this measure increases the plasma moves away from LTE. Note that $M_{KP}$ is calculated on the original instrument grid, while the enstrophy $\Omega$ is calculated on the Hermite grid.

Field-based proxies generally estimate the energy available for energy conversion. 
Here, we consider the Local Energy Transfer ($LET$) \citep{Sorriso-Valvo2015,Sorriso-Valvo2018,Sorriso2019PhRvL.122c5102S, Marino_Sorriso2023PhR..1006....1M}, which provides an approximated estimate of the local (in time $t$ and scale $\Delta t$) turbulent energy transfer,
\begin{equation}
\epsilon^{\pm} (t, \Delta t)= -\frac{3}{4} \frac{\Delta v_l(|\Delta \textbf{v}|^2+|\Delta \textbf{b}|^2)-2 \Delta b_l(\Delta \textbf{v} \cdot \Delta \textbf{b})}{\Delta t \langle v\rangle}   , 
\end{equation}
where $\Delta v_l$, $\Delta \textbf{v}$, $\Delta \textbf{b}$ are, respectively, the longitudinal velocity increments, the velocity and magnetic field increments computed with a lag $\Delta t$ equal to the SPAN-i cadence. 
Moreover, we adopt the partial variance of increments estimating the strength of magnetic field gradients \citep{Greco2008GeoRL}
\begin{equation}
    {\rm PVI} = \frac{|\Delta \textbf{b} |^2}{\langle|\Delta \textbf{b}|^2\rangle} ,
\end{equation}
where the increments are computed with a lag equal to the SPAN-i cadence, but on the much higher cadence MAG data. The lag choice is motivated by the idea of having for each VDF an associated value of the $LET$ and to have a measure of the field gradients, through the $\rm PVI$, on the same time scale at which the VDF is sampled. For both PVI and LET we use the standard 2-point increments, since at the scale of interest is at the bottom of the Kolmogorov range, where the spectra are not as steep as to require the more refined 5-point technique\citep{Cerri2019FrASS...6...64C}.

Fig.~\ref{fig:NoHornet} shows these quantities for the \textit{wave stream} (top) and the \textit{turbulent stream} (bottom). 
A good qualitative agreement between $M_{KP}$ and $\Omega/n^2$ (enstrophy normalized to the average stream density) is found for both streams (top panels), even though the two quantities are computed on different grids. This again supports the robustness of the analysis. Indeed, while the two measures are not formally identical, they both quantify deviations from Maxwellianity \citep{Pezzi2021MNRAS}.
The cause of the better agreement for the \textit{wave stream} with respect to \textit{turbulent stream} remains to be clarified.

Establishing quantitative correlations between energy conversion and dissipation variables in turbulent space plasmas is always complicated, since regional (adjacent peaks) rather than point-wise correlations naturally arise \citep{Servidio2017PhRvL.119t5101S, Yang2019MNRAS.482.4933Y, Yordanova2021ApJ...921...65Y}. In the two intervals analyzed here, we observe few instances in which the peaks or dips of PVI and enstrophy are simultaneous (the shaded areas numbered 3 and 5 in Fig. \ref{fig:NoHornet}, with dotted lines to aid the eye), while the correspondence is generally more regional (as in the shaded area 2). 
Cases with no evident correlations also appear (e.g., shaded area 4). 

A different argument holds for the $LET$ variable. Indeed, $LET$ peaks are much higher for the \textit{turbulent stream}, as expected for a stream with perpendicular sampling and closer to the Sun. 
In both intervals, most of the $LET$ peaks --indicative of enhanced turbulent energy transfer-- are associated with peaks in $M_{KP}$ and $\Omega$ (shaded areas 1, 3, 5, 6, 7). A similar behavior has been previously observed in Kelvin-Helmoltz vortices at the Earth magnetopause boundary layer \citep{Sorriso2019PhRvL.122c5102S}, where large values of the $LET$ were found to be associated with either beams or broad energization in the VDFs, but not with more Maxwellian VDFs.

These results suggest that an enhanced nonlinear transfer in the real space affects the phase space too, in agreement with the concept of a dual real-velocity space cascade.

Finally we observe that the Hermite spectrogram (bottom panels of Fig~\ref{fig:NoHornet}) is more disturbed and rapidly fluctuating for the \textit{turbulent stream} (confirming that the velocity-space cascade is intermittent \citep{Pezzi2018PhPl...25f0704P}) while for the \textit{wave stream} it is smoother and dominated by lower-$m$ modes.

\section{Discussion and conclusion} \label{sec:Discussion}
In this letter, we investigate the Hermite decomposition of the gyrotropized Parker Solar Probe ion velocity distribution functions for two different streams, a super-Alfvénic stream at $28 \ R_\odot$ and a sub-Alfvénic $11 \ R_\odot$ , referred to respectively as the \textit{wave 
stream} and \textit{turbulent stream}.

The streams under exam present similar level of fluctuations with respect to the background flow, a significant level of correlations between the velocity and magnetic field (high cross-helicity) but different $\beta_p$, angle to flow and velocity distribution functions shapes. The stream at $R \approx 28 R_{sun}$ presents coherent wave activity around the ion-cyclotron frequency and hammerhead distributions \citep{Verniero2020ApJS, Verniero2022ApJ}, while in the stream at $R \approx 11 R_{sun}$, where the sampling is rather perpendicular ($\theta_{BV} =116 ^\circ \pm 7^\circ $), the hammerhead features are not present.

These differences have a significant impact on the average Hermite spectrum and spectrogram. For the \textit{wave stream} the Hermite spectrum is dominated by the first few modes, while for the \textit{turbulent stream}, closer in, the higher order modes have relatively more power and the Hermite spectrogram is more intermittent. We speculate that this effect is related to (i) the presence of stronger turbulent fluctuations that, in turn, transfer more power to higher Hermite modes and, (ii) to the different sampling direction. The precise details of this mechanism must be investigated further. However, numerical efforts have provided compelling evidence for the ability of turbulence to transfer power to higher Hermite modes \citep{Pezzi2018PhPl...25f0704P}.  

Furthermore, we study the behavior of different energization and dissipation proxies for the two streams. Although, at variance with previous works \citep{Servidio2017PhRvL.119t5101S, Pezzi2018PhPl...25f0704P}, we do not observe a significant correlation between the enstrophy and the partial variance of increments of the magnetic field, our comparison of field-based (partial variance of increments and local energy transfer) and VDF-based (Kauffman-Paterson and the enstrophy) diagnostics reveals that there is good qualitative correlation between the peaks in local energy transfer and the ones in the Enstrophy and in the Kauffman-Paterson measures. These results support the idea that the turbulent cascade in nearly-reversible space and astrophysical plasmas is a \textit{dual} cascade, influencing both real and velocity space \citep{Shecko2008PPCF...50l4024S, Servidio2017PhRvL.119t5101S, Cerri2018ApJ, Pezzi2018PhPl...25f0704P} and are important with respect to the long-standing problem of how heating and dissipation happen in collisionless plasma. 
Indeed, the preliminary evidence of the presence of a phase-space cascade in the inner heliosphere corroborates the argument that fine velocity distribution features can be crucial in irreversibly dissipating turbulent energy even when the collisional frequency is small \citep{Shecko2008PPCF...50l4024S, 2016PhRvL.116n5001P}. 

Future efforts will complement this analysis adopting different in-situ missions, such as Solar Orbiter, and in the context of existent and future multi-spacecraft missions (MMS, Helioswarm, and Plasma Observatory). Novel methods to reconstruct the velocity distribution functions, such as the one based on the Slepian function \citep{SrijanMicheal2025ApJ...982...96B}, will also be considered since they have the potential to improve the study of the phase-space cascade. 

\begin{acknowledgments}
We thank C. H. K. Chen, M. Terres, J. Coburn, D. Manzini and S.S. Cerri for useful discussions. 
This research was preliminary discussed within (a) the ISSI International Team Project 23‐588 (“Unveiling Energy Conversion and Dissipation in Non‐Equilibrium Space Plasmas”), and (b) the International Exchanges Cost Share scheme/Joint Bilateral Agreement project “Multi-scale electrostatic energisation of plasmas: comparison of collective processes in laboratory and space” funded by the Royal Society (UK) and the Consiglio Nazionale delle Ricerche (Italy) (award numbers IEC\textbackslash R2\textbackslash222050 and SAC.AD002.043.021).
OP, DP, FV, LSV, and FP acknowledge the support of the PRIN 2022 project ``2022KL38BK - The ULtimate fate of TuRbulence from space to laboratory plAsmas (ULTRA)'' (Master CUP B53D23004850006) by the Italian Ministry of University and Research, funded under the National Recovery and Resilience Plan (NRRP), Mission 4 – Component C2 – Investment 1.1, ``Fondo per il Programma Nazionale di Ricerca e Progetti di Rilevante Interesse Nazionale (PRIN 2022)'' (PE9) by the European Union – NextGenerationEU. 
LSV was supported by the Swedish Research Council (VR) Research Grant N. 2022-03352.
AL, LSV and FP acknowledge the project ``Data-based predictions of solar energetic particle arrival to the Earth: ensuring space data and technology integrity from hazardous solar activity events'' (CUP H53D23011020001) `Finanziato dall'Unione europea - Next Generation EU' PIANO NAZIONALE DI RIPRESA E RESILIENZA (PNRR) Missione 4 ``Istruzione e Ricerca'' - Componente C2 Investimento 1.1, ``Fondo per il Programma Nazionale di Ricerca e Progetti di Rilevante Interesse Nazionale (PRIN)'' Settore PE09. 

FP and AL acknowledges support from the Research Foundation – Flanders (FWO) Junior research project on fundamental research G020224N.

FV received funding from the European Union’s Horizon Europe research and innovation program under Grant Agreement No. 101082633421 - AutomaticS in spAce exPloration (ASAP).

S. S. acknowledges the Space It Up project
funded by the Italian Space Agency, ASI, and the Ministry of Uni-
versity and Research, MUR, under contract n. 2024-5-E.0 - CUP n.
I53D24000060005
\end{acknowledgments}

\software{MHDTurbPy (see the function TurbPy.flucts to compute increments and  TurbPy.remove\_wheel\_noise to remove the reaction wheel peaks from the magnetic field spectrum}) 
\citep{nikos_sioulas_2023_7572468},  
          PySpedas \textbf{\citep{Angelopulos2019SSRv..215....9A}},
          PyUltra (https://github.com/orestepezzi/),
          PyPesto (currently under development)
          .

\appendix
\section{Gyrotropization of the velocity distribution function}\label{app:gyro}

\begin{figure}[htb!]
\includegraphics[width=1\textwidth]{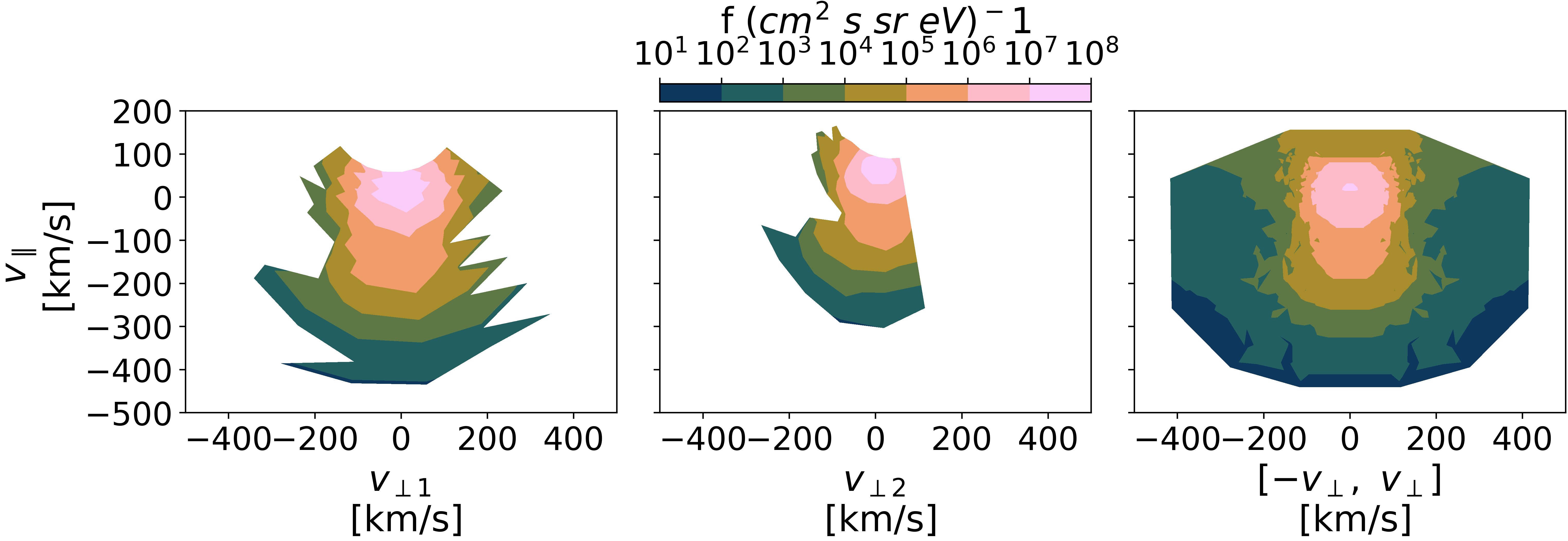}
    \caption{VDF processing for the \textit{wave stream}. VDF averaged over $\phi$ and plotted in the Energy-$\theta$ plane (left), VDF averaged over $\theta$ and plotted in the Energy-$\phi$ plane (center), VDF after the procedure describer in Sec~\ref{sec:data_methods} before the interpolation into the Hermite Grid. 
    \label{fig:fig_vdfs_appendix}}
\end{figure}

The aim of this section is to give more details on the preprocessing and gyrotropization of VDFs briefly discussed in Sec~\ref{sec:data_methods}. In Fig~\ref{fig:fig_vdfs_appendix} we plot the same distribution function of Fig~\ref{fig:vdfs} (top) in FAC integrated along $\phi$ and plotted in the Energy-$\theta$ plane (left), integrated along $\theta$ and plotted in the Energy-$\phi$ plane (center) and its gyrotropized version following \cite{Bowen2022PhRvL}.
$\theta$ and $\phi$ are respectevely the elevation and azimuthal angle of SPANi \citep{Livi2022ApJ}. The VDFs are centered by subtracting the bulk speed.

The limited field of view due to the heat shield is visible in the Energy-$\phi$ plane (center).
The rightmost panel shows the full VDF (not integrated along any direction) collapsed in the new grid where $v_\parallel$ is unchanged and $v_\perp = \sqrt{v_{\perp1}^2 + v_{\perp2}^2}$, extended to the $-v_\perp$ by imposing $f(-v_\perp) = f(v_\perp)$. The Hermite transform is performed on the latter VDF, once the grid has been normalized to the parallel and perpendicular thermal speeds and interpolated onto the Hermite grid.

This procedure provides a reduced distribution function that accurately represents the PSP measurements while mitigating the effects of FOV limitations.

\section{Hermite spectra: data vs bi-Maxwellians}\label{app:biMax}

\begin{figure}[htb!]
\centering
\includegraphics[width=0.7\textwidth]{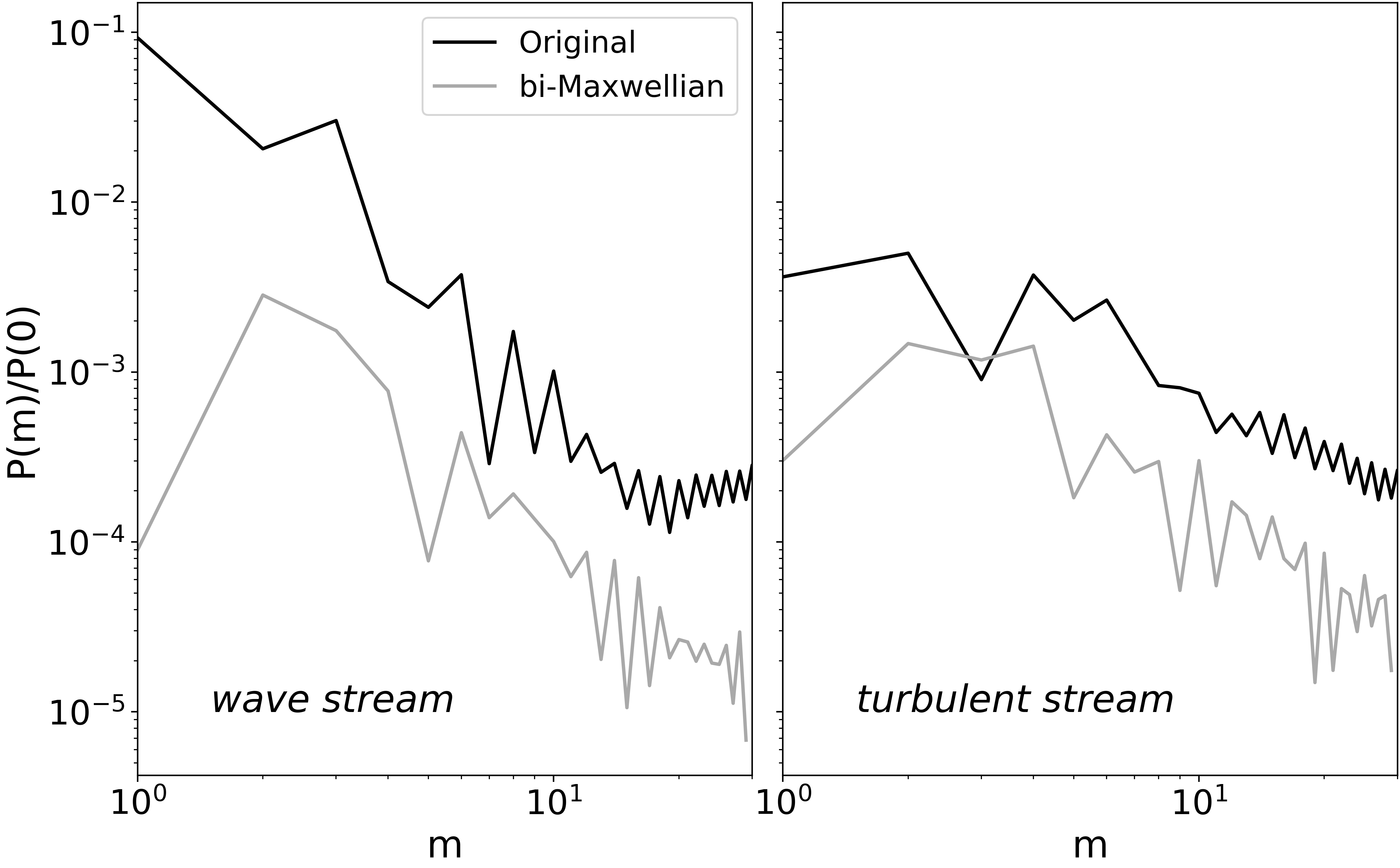}
    \caption{Original Hermite spectrum for the \textit{wave stream} (left) and \textit{turbulent stream} (right) in black compared with the Hermite spectrum of the underlying bi-Maxwellian in gray.}
    \label{fig:spectra_biMax}
\end{figure}

In order to evaluate whether the Hermite spectra shown in Fig~\ref{fig:spectral_prop} are meaningful we perform the same analysis on bi-Maxwellian distributions obtained from the density and parallel and perpendicular temperature of each measurements in the two streams. For each bi-Maxwellian, we compute the Hermite spectrum and we then average to obtain the spectra shown in Fig~\ref{fig:spectra_biMax}.

The purpose of the comparison in Fig.~\ref{fig:spectra_biMax} is to demonstrate that the measured spectra are meaningful: the observed signal lies above that of the underlying bi-Maxwellians, which can be considered the lowest-order non-trivial VDF against which to benchmark our spectra. The spectra are also above the one count noise level (not shown).

\newpage

\bibliography{sample631}{}

@ARTICLE{2015JFM...770R...1M,
       author = {{Meyrand}, R. and {Kiyani}, K.~H. and {Galtier}, S.},
        title = "{Weak magnetohydrodynamic turbulence and intermittency}",
      journal = {Journal of Fluid Mechanics},
         year = 2015,
        month = may,
       volume = {770},
          eid = {R1},
        pages = {R1},
          doi = {10.1017/jfm.2015.141},
       adsurl = {https://ui.adsabs.harvard.edu/abs/2015JFM...770R...1M}
}

@ARTICLE{2022JPlPh..88e1501S,
       author = {{Schekochihin}, Alexander A.},
        title = "{MHD turbulence: a biased review}",
      journal = {Journal of Plasma Physics},
         year = 2022,
        month = oct,
       volume = {88},
       number = {5},
          eid = {155880501},
        pages = {155880501},
          doi = {10.1017/S0022377822000721},
archivePrefix = {arXiv},
       eprint = {2010.00699},
 primaryClass = {physics.plasm-ph},
       adsurl = {https://ui.adsabs.harvard.edu/abs/2022JPlPh..88e1501S}
}

@ARTICLE{Boldyrev2006PhRvL..96k5002B,
       author = {{Boldyrev}, Stanislav},
        title = "{Spectrum of Magnetohydrodynamic Turbulence}",
      journal = {\prl},
         year = 2006,
        month = mar,
       volume = {96},
       number = {11},
          eid = {115002},
        pages = {115002},
          doi = {10.1103/PhysRevLett.96.115002},
archivePrefix = {arXiv},
       eprint = {astro-ph/0511290},
 primaryClass = {astro-ph},
       adsurl = {https://ui.adsabs.harvard.edu/abs/2006PhRvL..96k5002B}
}

@ARTICLE{Boldyrev2005ApJ...626L..37B,
       author = {{Boldyrev}, Stanislav},
        title = "{On the Spectrum of Magnetohydrodynamic Turbulence}",
      journal = {\apjl},
         year = 2005,
        month = jun,
       volume = {626},
       number = {1},
        pages = {L37-L40},
          doi = {10.1086/431649},
archivePrefix = {arXiv},
       eprint = {astro-ph/0503053},
 primaryClass = {astro-ph},
       adsurl = {https://ui.adsabs.harvard.edu/abs/2005ApJ...626L..37B}
}

@article{Mondal2025,
doi = {10.3847/1538-4357/adba54},
url = {https://doi.org/10.3847/1538-4357/adba54},
year = {2025},
month = {mar},
publisher = {The American Astronomical Society},
volume = {982},
number = {2},
pages = {199},
author = {Mondal, Shiladittya and Banerjee, Supratik and Sorriso-Valvo, Luca},
title = {Emergence of Two Inertial Subranges in Solar Wind Turbulence: Dependence on Heliospheric Distance and Solar Activity},
journal = {The Astrophysical Journal}
}

@ARTICLE{CaroneF2021A&A...656A..16C,
       author = {{Carbone}, F. and {Sorriso-Valvo}, L. and {Khotyaintsev}, Yu. V. and {Steinvall}, K. and {Vecchio}, A. and {Telloni}, D. and {Yordanova}, E. and {Graham}, D.~B. and {Edberg}, N.~J.~T. and {Eriksson}, A.~I. and {Johansson}, E.~P.~G. and {V{\'a}sconez}, C.~L. and {Maksimovic}, M. and {Bruno}, R. and {D'Amicis}, R. and {Bale}, S.~D. and {Chust}, T. and {Krasnoselskikh}, V. and {Kretzschmar}, M. and {Lorf{\`e}vre}, E. and {Plettemeier}, D. and {Sou{\v{c}}ek}, J. and {Steller}, M. and {{\v{S}}tver{\'a}k}, {\v{S}}. and {Tr{\'a}vn{\'\i}{\v{c}}ek}, P. and {Vaivads}, A. and {Horbury}, T.~S. and {O'Brien}, H. and {Angelini}, V. and {Evans}, V.},
        title = "{Statistical study of electron density turbulence and ion-cyclotron waves in the inner heliosphere: Solar Orbiter observations}",
      journal = {\aap},
         year = 2021,
        month = dec,
       volume = {656},
          eid = {A16},
        pages = {A16},
          doi = {10.1051/0004-6361/202140931},
archivePrefix = {arXiv},
       eprint = {2105.07790},
 primaryClass = {physics.space-ph},
       adsurl = {https://ui.adsabs.harvard.edu/abs/2021A&A...656A..16C}
}

@ARTICLE{Klein_2015ApJ...801L..18K,
       author = {{Klein}, Kristopher G. and {Perez}, Jean C. and {Verscharen}, Daniel and {Mallet}, Alfred and {Chandran}, Benjamin D.~G.},
        title = "{A Modified Version of Taylor{\textquoteright}s Hypothesis for Solar Probe Plus Observations}",
      journal = {\apjl},
         year = 2015,
        month = mar,
       volume = {801},
       number = {1},
          eid = {L18},
        pages = {L18},
          doi = {10.1088/2041-8205/801/1/L18},
archivePrefix = {arXiv},
       eprint = {1412.3786},
 primaryClass = {physics.space-ph},
       adsurl = {https://ui.adsabs.harvard.edu/abs/2015ApJ...801L..18K}
}

@ARTICLE{OwensFors2013LRSP...10....5O,
       author = {{Owens}, Mathew J. and {Forsyth}, Robert J.},
        title = "{The Heliospheric Magnetic Field}",
      journal = {Living Reviews in Solar Physics},
         year = 2013,
        month = dec,
       volume = {10},
       number = {1},
          eid = {5},
        pages = {5},
          doi = {10.12942/lrsp-2013-5},
       adsurl = {https://ui.adsabs.harvard.edu/abs/2013LRSP...10....5O}
}

@ARTICLE{Angelopulos2019SSRv..215....9A,
       author = {{Angelopoulos}, V. and {Cruce}, P. and {Drozdov}, A. and {Grimes}, E.~W. and {Hatzigeorgiu}, N. and {King}, D.~A. and {Larson}, D. and {Lewis}, J.~W. and {McTiernan}, J.~M. and {Roberts}, D.~A. and {Russell}, C.~L. and {Hori}, T. and {Kasahara}, Y. and {Kumamoto}, A. and {Matsuoka}, A. and {Miyashita}, Y. and {Miyoshi}, Y. and {Shinohara}, I. and {Teramoto}, M. and {Faden}, J.~B. and {Halford}, A.~J. and {McCarthy}, M. and {Millan}, R.~M. and {Sample}, J.~G. and {Smith}, D.~M. and {Woodger}, L.~A. and {Masson}, A. and {Narock}, A.~A. and {Asamura}, K. and {Chang}, T.~F. and {Chiang}, C. -Y. and {Kazama}, Y. and {Keika}, K. and {Matsuda}, S. and {Segawa}, T. and {Seki}, K. and {Shoji}, M. and {Tam}, S.~W.~Y. and {Umemura}, N. and {Wang}, B. -J. and {Wang}, S. -Y. and {Redmon}, R. and {Rodriguez}, J.~V. and {Singer}, H.~J. and {Vandegriff}, J. and {Abe}, S. and {Nose}, M. and {Shinbori}, A. and {Tanaka}, Y. -M. and {UeNo}, S. and {Andersson}, L. and {Dunn}, P. and {Fowler}, C. and {Halekas}, J.~S. and {Hara}, T. and {Harada}, Y. and {Lee}, C.~O. and {Lillis}, R. and {Mitchell}, D.~L. and {Argall}, M.~R. and {Bromund}, K. and {Burch}, J.~L. and {Cohen}, I.~J. and {Galloy}, M. and {Giles}, B. and {Jaynes}, A.~N. and {Le Contel}, O. and {Oka}, M. and {Phan}, T.~D. and {Walsh}, B.~M. and {Westlake}, J. and {Wilder}, F.~D. and {Bale}, S.~D. and {Livi}, R. and {Pulupa}, M. and {Whittlesey}, P. and {DeWolfe}, A. and {Harter}, B. and {Lucas}, E. and {Auster}, U. and {Bonnell}, J.~W. and {Cully}, C.~M. and {Donovan}, E. and {Ergun}, R.~E. and {Frey}, H.~U. and {Jackel}, B. and {Keiling}, A. and {Korth}, H. and {McFadden}, J.~P. and {Nishimura}, Y. and {Plaschke}, F. and {Robert}, P. and {Turner}, D.~L. and {Weygand}, J.~M. and {Candey}, R.~M. and {Johnson}, R.~C. and {Kovalick}, T. and {Liu}, M.~H. and {McGuire}, R.~E. and {Breneman}, A. and {Kersten}, K. and {Schroeder}, P.},
        title = "{The Space Physics Environment Data Analysis System (SPEDAS)}",
      journal = {\ssr},
         year = 2019,
        month = jan,
       volume = {215},
       number = {1},
          eid = {9},
        pages = {9},
          doi = {10.1007/s11214-018-0576-4},
       adsurl = {https://ui.adsabs.harvard.edu/abs/2019SSRv..215....9A}
}

@ARTICLE{Nour2023SSRv..219....8R,
       author = {{Raouafi}, N.~E. and {Matteini}, L. and {Squire}, J. and {Badman}, S.~T. and {Velli}, M. and {Klein}, K.~G. and {Chen}, C.~H.~K. and {Matthaeus}, W.~H. and {Szabo}, A. and {Linton}, M. and {Allen}, R.~C. and {Szalay}, J.~R. and {Bruno}, R. and {Decker}, R.~B. and {Akhavan-Tafti}, M. and {Agapitov}, O.~V. and {Bale}, S.~D. and {Bandyopadhyay}, R. and {Battams}, K. and {Ber{\v{c}}i{\v{c}}}, L. and {Bourouaine}, S. and {Bowen}, T.~A. and {Cattell}, C. and {Chandran}, B.~D.~G. and {Chhiber}, R. and {Cohen}, C.~M.~S. and {D'Amicis}, R. and {Giacalone}, J. and {Hess}, P. and {Howard}, R.~A. and {Horbury}, T.~S. and {Jagarlamudi}, V.~K. and {Joyce}, C.~J. and {Kasper}, J.~C. and {Kinnison}, J. and {Laker}, R. and {Liewer}, P. and {Malaspina}, D.~M. and {Mann}, I. and {McComas}, D.~J. and {Niembro-Hernandez}, T. and {Nieves-Chinchilla}, T. and {Panasenco}, O. and {Pokorn{\'y}}, P. and {Pusack}, A. and {Pulupa}, M. and {Perez}, J.~C. and {Riley}, P. and {Rouillard}, A.~P. and {Shi}, C. and {Stenborg}, G. and {Tenerani}, A. and {Verniero}, J.~L. and {Viall}, N. and {Vourlidas}, A. and {Wood}, B.~E. and {Woodham}, L.~D. and {Woolley}, T.},
        title = "{Parker Solar Probe: Four Years of Discoveries at Solar Cycle Minimum}",
      journal = {\ssr},
         year = 2023,
        month = feb,
       volume = {219},
       number = {1},
          eid = {8},
        pages = {8},
          doi = {10.1007/s11214-023-00952-4},
archivePrefix = {arXiv},
       eprint = {2301.02727},
 primaryClass = {astro-ph.SR},
       adsurl = {https://ui.adsabs.harvard.edu/abs/2023SSRv..219....8R}
}

@ARTICLE{Matthaeus1982,
       author = {{Matthaeus}, W.~H. and {Goldstein}, M.~L. and {Smith}, C.},
        title = "{Evaluation of magnetic helicity in homogeneous turbulence}",
      journal = {\prl},
         year = 1982,
        month = may,
       volume = {48},
       number = {18},
        pages = {1256-1259},
          doi = {10.1103/PhysRevLett.48.1256}
}

@ARTICLE{Yang2019MNRAS.482.4933Y,
       author = {{Yang}, Yan and {Wan}, Minping and {Matthaeus}, William H. and {Sorriso-Valvo}, Luca and {Parashar}, Tulasi N. and {Lu}, Quanming and {Shi}, Yipeng and {Chen}, Shiyi},
        title = "{Scale dependence of energy transfer in turbulent plasma}",
      journal = {\mnras},
         year = 2019,
        month = feb,
       volume = {482},
       number = {4},
        pages = {4933-4940},
          doi = {10.1093/mnras/sty2977},
archivePrefix = {arXiv},
       eprint = {1809.05677},
 primaryClass = {physics.space-ph},
       adsurl = {https://ui.adsabs.harvard.edu/abs/2019MNRAS.482.4933Y}
}

@ARTICLE{Nikos2024arXiv240404055S,
       author = {{Sioulas}, Nikos and {Zikopoulos}, Themistocles and {Shi}, Chen and {Velli}, Marco and {Bowen}, Trevor and {Mallet}, Alfred and {Sorriso-Valvo}, Luca and {Verdini}, Andrea and {Chandran}, B.~D.~G. and {Martinovi{\'c}}, Mihailo M. and {Cerri}, S.~S. and {Davis}, Nooshin and {Dunn}, Corina},
        title = "{Higher-Order Analysis of Three-Dimensional Anisotropy in Imbalanced Alfv{\'e}nic Turbulence}",
      journal = {arXiv e-prints},
         year = 2024,
        month = apr,
          eid = {arXiv:2404.04055},
        pages = {arXiv:2404.04055},
          doi = {10.48550/arXiv.2404.04055},
archivePrefix = {arXiv},
       eprint = {2404.04055},
 primaryClass = {physics.space-ph},
       adsurl = {https://ui.adsabs.harvard.edu/abs/2024arXiv240404055S}
}

@ARTICLE{Cho2019ApJ...874...75C,
       author = {{Cho}, Jungyeon},
        title = "{A Technique for Removing Large-scale Variations in Regularly and Irregularly Spaced Data}",
      journal = {\apj},
         year = 2019,
        month = mar,
       volume = {874},
       number = {1},
          eid = {75},
        pages = {75},
          doi = {10.3847/1538-4357/ab06f3},
       adsurl = {https://ui.adsabs.harvard.edu/abs/2019ApJ...874...75C}
}

@ARTICLE{Cerri2019FrASS...6...64C,
       author = {{Cerri}, Silvio Sergio and {Gro{\v{s}}elj}, Daniel and {Franci}, Luca},
        title = "{Kinetic plasma turbulence: recent insights and open questions from 3D3V simulations}",
      journal = {Frontiers in Astronomy and Space Sciences},
         year = 2019,
        month = oct,
       volume = {6},
          eid = {64},
        pages = {64},
          doi = {10.3389/fspas.2019.00064},
archivePrefix = {arXiv},
       eprint = {1909.11525},
 primaryClass = {astro-ph.SR},
       adsurl = {https://ui.adsabs.harvard.edu/abs/2019FrASS...6...64C}
}

@ARTICLE{Chen2011MNRAS.415.3219C,
       author = {{Chen}, C.~H.~K. and {Mallet}, A. and {Yousef}, T.~A. and {Schekochihin}, A.~A. and {Horbury}, T.~S.},
        title = "{Anisotropy of Alfv{\'e}nic turbulence in the solar wind and numerical simulations}",
      journal = {\mnras},
         year = 2011,
        month = aug,
       volume = {415},
       number = {4},
        pages = {3219-3226},
          doi = {10.1111/j.1365-2966.2011.18933.x},
archivePrefix = {arXiv},
       eprint = {1009.0662},
 primaryClass = {physics.space-ph},
       adsurl = {https://ui.adsabs.harvard.edu/abs/2011MNRAS.415.3219C}
}

@ARTICLE{Shebalin1983JPlPh..29..525S,
       author = {{Shebalin}, J.~V. and {Matthaeus}, W.~H. and {Montgomery}, D.},
        title = "{Anisotropy in MHD turbulence due to a mean magnetic field}",
      journal = {Journal of Plasma Physics},
         year = 1983,
        month = jun,
       volume = {29},
       number = {3},
        pages = {525-547},
          doi = {10.1017/S0022377800000933},
       adsurl = {https://ui.adsabs.harvard.edu/abs/1983JPlPh..29..525S}
}

@ARTICLE{Jannet2021JGRA..12628543J,
       author = {{Jannet}, G. and {Dudok de Wit}, T. and {Krasnoselskikh}, V. and {Kretzschmar}, M. and {Fergeau}, P. and {Bergerard-Timofeeva}, M. and {Agrapart}, C. and {Brochot}, J. -Y. and {Chalumeau}, G. and {Martin}, P. and {Revillet}, C. and {Bale}, S.~D. and {Maksimovic}, M. and {Bowen}, T.~A. and {Brysbaert}, C. and {Goetz}, K. and {Guilhem}, E. and {Harvey}, P.~R. and {Leray}, V. and {Lorf{\`e}vre}, E.},
        title = "{Measurement of Magnetic Field Fluctuations in the Parker Solar Probe and Solar Orbiter Missions}",
      journal = {Journal of Geophysical Research (Space Physics)},
         year = 2021,
        month = feb,
       volume = {126},
       number = {2},
          eid = {e28543},
        pages = {e28543},
          doi = {10.1029/2020JA028543},
       adsurl = {https://ui.adsabs.harvard.edu/abs/2021JGRA..12628543J}
}

@ARTICLE{Thierry2022JGRA..12730018D,
       author = {{Dudok de Wit}, T. and {Krasnoselskikh}, V.~V. and {Agapitov}, O. and {Froment}, C. and {Larosa}, A. and {Bale}, S.~D. and {Bowen}, T. and {Goetz}, K. and {Harvey}, P. and {Jannet}, G. and {Kretzschmar}, M. and {MacDowall}, R.~J. and {Malaspina}, D. and {Martin}, P. and {Page}, B. and {Pulupa}, M. and {Revillet}, C.},
        title = "{First Results From the SCM Search-Coil Magnetometer on Parker Solar Probe}",
      journal = {Journal of Geophysical Research (Space Physics)},
         year = 2022,
        month = apr,
       volume = {127},
       number = {4},
          eid = {e30018},
        pages = {e30018},
          doi = {10.1029/2021JA030018},
       adsurl = {https://ui.adsabs.harvard.edu/abs/2022JGRA..12730018D}
}

@ARTICLE{Yordanova2021ApJ...921...65Y,
       author = {{Yordanova}, Emiliya and {V{\"o}r{\"o}s}, Zolt{\'a}n and {Sorriso-Valvo}, Luca and {Dimmock}, Andrew P. and {Kilpua}, Emilia},
        title = "{A Possible Link between Turbulence and Plasma Heating}",
      journal = {\apj},
         year = 2021,
        month = nov,
       volume = {921},
       number = {1},
          eid = {65},
        pages = {65},
          doi = {10.3847/1538-4357/ac1942},
archivePrefix = {arXiv},
       eprint = {2108.01376},
 primaryClass = {physics.plasm-ph},
       adsurl = {https://ui.adsabs.harvard.edu/abs/2021ApJ...921...65Y}
}

@ARTICLE{Pecora2021A&A...650A..20P,
       author = {{Pecora}, F. and {Servidio}, S. and {Greco}, A. and {Matthaeus}, W.~H.},
        title = "{Identification of coherent structures in space plasmas: the magnetic helicity-PVI method}",
      journal = {\aap},
         year = 2021,
        month = jun,
       volume = {650},
          eid = {A20},
        pages = {A20},
          doi = {10.1051/0004-6361/202039639},
archivePrefix = {arXiv},
       eprint = {2010.09500},
 primaryClass = {physics.plasm-ph},
       adsurl = {https://ui.adsabs.harvard.edu/abs/2021A&A...650A..20P}
}

@ARTICLE{bruno2013LRSP...10....2B,
       author = {{Bruno}, Roberto and {Carbone}, Vincenzo},
        title = "{The Solar Wind as a Turbulence Laboratory}",
      journal = {Living Reviews in Solar Physics},
         year = 2013,
        month = dec,
       volume = {10},
       number = {1},
          eid = {2},
        pages = {2},
          doi = {10.12942/lrsp-2013-2},
       adsurl = {https://ui.adsabs.harvard.edu/abs/2013LRSP...10....2B}
}

@article{Sorriso-Valvo2015,
	doi = {10.1088/0004-637x/807/1/86},
	url = {https://doi.org/10.1088/0004-637x/807/1/86},
	year = 2015,
	month = {jul},
	publisher = {American Astronomical Society},
	volume = {807},
	number = {1},
	pages = {86},
	author = {L. Sorriso-Valvo and R. Marino and L. Lijoi and S. Perri and V. Carbone},
	title = {{SELF}-{CONSISTENT} {CASTAING} {DISTRIBUTION} {OF} {SOLAR} {WIND} {TURBULENT} {FLUCTUATIONS}},
	journal = {\apj}
}

@ARTICLE{Sorriso-Valvo2018,
       author = {{Sorriso-Valvo}, Luca and {Carbone}, Francesco and {Perri}, Silvia and {Greco}, Antonella and {Marino}, Raffaele and {Bruno}, Roberto},
        title = "{On the Statistical Properties of Turbulent Energy Transfer Rate in the Inner Heliosphere}",
      journal = {\solphys},
         year = 2018,
        month = jan,
       volume = {293},
       number = {1},
          eid = {10},
        pages = {10},
          doi = {10.1007/s11207-017-1229-6},
archivePrefix = {arXiv},
       eprint = {1712.09825},
 primaryClass = {physics.space-ph},
       adsurl = {https://ui.adsabs.harvard.edu/abs/2018SoPh..293...10S}
}

@ARTICLE{Shecko2016JPlPh..82b9012S,
       author = {{Schekochihin}, A.~A. and {Parker}, J.~T. and {Highcock}, E.~G. and {Dellar}, P.~J. and {Dorland}, W. and {Hammett}, G.~W.},
        title = "{Phase mixing versus nonlinear advection in drift-kinetic plasma turbulence}",
      journal = {Journal of Plasma Physics},
         year = 2016,
        month = apr,
       volume = {82},
       number = {2},
          eid = {905820212},
        pages = {905820212},
          doi = {10.1017/S0022377816000374},
archivePrefix = {arXiv},
       eprint = {1508.05988},
 primaryClass = {physics.plasm-ph},
       adsurl = {https://ui.adsabs.harvard.edu/abs/2016JPlPh..82b9012S}
}

@ARTICLE{cassak2023,
       author = {{Cassak}, Paul A. and {Barbhuiya}, M. Hasan and {Liang}, Haoming and {Argall}, Matthew R.},
        title = "{Quantifying Energy Conversion in Higher-Order Phase Space Density Moments in Plasmas}",
      journal = {\prl},
         year = 2023,
        month = feb,
       volume = {130},
       number = {8},
          eid = {085201},
        pages = {085201},
          doi = {10.1103/PhysRevLett.130.085201},
archivePrefix = {arXiv},
       eprint = {2306.01106},
 primaryClass = {physics.plasm-ph},
       adsurl = {https://ui.adsabs.harvard.edu/abs/2023PhRvL.130h5201C}
}

@ARTICLE{2016PhRvL.116n5001P,
       author = {{Pezzi}, Oreste and {Valentini}, Francesco and {Veltri}, Pierluigi},
        title = "{Collisional Relaxation of Fine Velocity Structures in Plasmas}",
      journal = {\prl},
         year = 2016,
        month = apr,
       volume = {116},
       number = {14},
          eid = {145001},
        pages = {145001},
          doi = {10.1103/PhysRevLett.116.145001},
       adsurl = {https://ui.adsabs.harvard.edu/abs/2016PhRvL.116n5001P}
}

@article{Sorriso-Valvo2023,
	author = {Sorriso-Valvo, L. and {Marino, R.} and {Foldes, R.} and {L\'ev\^eque, E.} and {D\'{}Amicis, R.} and {Bruno, R.} and {Telloni, D.} and {Yordanova, E.}},
	title = {Helios 2 observations of solar wind turbulence decay in the inner heliosphere},
	DOI= "10.1051/0004-6361/202244889",
	url= "https://doi.org/10.1051/0004-6361/202244889",
	journal = {A\&A},
	year = 2023,
	volume = 672,
	pages = "A13",
}

@Article{Bruno2003,
author = {Bruno, R. and Carbone, V. and Sorriso-Valvo, L. and Bavassano, B.},
title = {Radial evolution of solar wind intermittency in the inner heliosphere},
journal = {J. of Geophys. Res.: Space Phys.},
volume = {108},
number = {A3},
issn = {2156-2202},
pages = {1130},
year = {2003}
}

@Book{Frisch1995,
	title = "Turbulence: The Legacy of A.N. Kolmogorov",
	author = "Uriel Frisch",
	publisher = "Cambridge University Press",
	year = "1995"
}

@software{nikos_sioulas_2023_7572468,
  author       = {Nikos Sioulas},
  title        = {MHDTurbPy},
  month        = jan,
  year         = 2023,
  publisher    = {Zenodo},
  version      = {0.1.0},
  doi          = {10.5281/zenodo.7572468},
  url          = {https://doi.org/10.5281/zenodo.7572468}
}

@Article{CrameriNatureComm2020,
author="Crameri, Fabio
and Shephard, Grace E.
and Heron, Philip J.",
title="The misuse of colour in science communication",
journal="Nature Communications",
year="2020",
month="Oct",
day="28",
volume="11",
number="1",
pages="5444",
issn="2041-1723",
doi="10.1038/s41467-020-19160-7",
url="https://doi.org/10.1038/s41467-020-19160-7"
}

@ARTICLE{Wu2023ApJ...951...98W,
       author = {{Wu}, Ziqi and {He}, Jiansen and {Duan}, Die and {Zhu}, Xingyu and {Hou}, Chuanpeng and {Verscharen}, Daniel and {Nicolaou}, Georgios and {Owen}, Christopher J. and {Fedorov}, Andrey and {Louarn}, Philippe},
        title = "{Ion Energization and Thermalization in Magnetic Reconnection Exhaust Region in the Solar Wind}",
      journal = {\apj},
         year = 2023,
        month = jul,
       volume = {951},
       number = {2},
          eid = {98},
        pages = {98},
          doi = {10.3847/1538-4357/accf9b},
       adsurl = {https://ui.adsabs.harvard.edu/abs/2023ApJ...951...98W}
}

@ARTICLE{Parker2016PhPl...23g0703P,
       author = {{Parker}, J.~T. and {Highcock}, E.~G. and {Schekochihin}, A.~A. and {Dellar}, P.~J.},
        title = "{Suppression of phase mixing in drift-kinetic plasma turbulence}",
      journal = {Physics of Plasmas},
         year = 2016,
        month = jul,
       volume = {23},
       number = {7},
          eid = {070703},
        pages = {070703},
          doi = {10.1063/1.4958954},
archivePrefix = {arXiv},
       eprint = {1603.06968},
 primaryClass = {physics.plasm-ph},
       adsurl = {https://ui.adsabs.harvard.edu/abs/2016PhPl...23g0703P}
}

@ARTICLE{Tatsuno2009PhRvL.103a5003T,
       author = {{Tatsuno}, T. and {Dorland}, W. and {Schekochihin}, A.~A. and {Plunk}, G.~G. and {Barnes}, M. and {Cowley}, S.~C. and {Howes}, G.~G.},
        title = "{Nonlinear Phase Mixing and Phase-Space Cascade of Entropy in Gyrokinetic Plasma Turbulence}",
      journal = {\prl},
         year = 2009,
        month = jul,
       volume = {103},
       number = {1},
          eid = {015003},
        pages = {015003},
          doi = {10.1103/PhysRevLett.103.015003},
archivePrefix = {arXiv},
       eprint = {0811.2538},
 primaryClass = {physics.plasm-ph},
       adsurl = {https://ui.adsabs.harvard.edu/abs/2009PhRvL.103a5003T}
}

@ARTICLE{HowesLab2018PhPl...25e5501H,
       author = {{Howes}, Gregory G.},
        title = "{Laboratory space physics: Investigating the physics of space plasmas in the laboratory}",
      journal = {Physics of Plasmas},
         year = 2018,
        month = may,
       volume = {25},
       number = {5},
          eid = {055501},
        pages = {055501},
          doi = {10.1063/1.5025421},
archivePrefix = {arXiv},
       eprint = {1802.04154},
 primaryClass = {physics.space-ph},
       adsurl = {https://ui.adsabs.harvard.edu/abs/2018PhPl...25e5501H}
}

@ARTICLE{HuangTTDvsLD2024JPlPh..90d5301H,
       author = {{Huang}, Rui and {Howes}, Gregory G. and {McCubbin}, Andrew J.},
        title = "{The velocity-space signature of transit-time damping}",
      journal = {Journal of Plasma Physics},
         year = 2024,
        month = sep,
       volume = {90},
       number = {4},
          eid = {535900401},
        pages = {535900401},
          doi = {10.1017/S0022377824000667},
archivePrefix = {arXiv},
       eprint = {2401.16697},
 primaryClass = {physics.plasm-ph},
       adsurl = {https://ui.adsabs.harvard.edu/abs/2024JPlPh..90d5301H}
}

@ARTICLE{Nastac2024PhRvE.109f5210N,
       author = {{Nastac}, Michael L. and {Ewart}, Robert J. and {Sengupta}, Wrick and {Schekochihin}, Alexander A. and {Barnes}, Michael and {Dorland}, William D.},
        title = "{Phase-space entropy cascade and irreversibility of stochastic heating in nearly collisionless plasma turbulence}",
      journal = {\pre},
         year = 2024,
        month = jun,
       volume = {109},
       number = {6},
          eid = {065210},
        pages = {065210},
          doi = {10.1103/PhysRevE.109.065210},
archivePrefix = {arXiv},
       eprint = {2310.18211},
 primaryClass = {physics.plasm-ph},
       adsurl = {https://ui.adsabs.harvard.edu/abs/2024PhRvE.109f5210N}
}

@ARTICLE{Marino_Sorriso2023PhR..1006....1M,
       author = {{Marino}, Raffaele and {Sorriso-Valvo}, Luca},
        title = "{Scaling laws for the energy transfer in space plasma turbulence}",
      journal = {\physrep},
         year = 2023,
        month = mar,
       volume = {1006},
        pages = {1-144},
          doi = {10.1016/j.physrep.2022.12.001},
       adsurl = {https://ui.adsabs.harvard.edu/abs/2023PhR..1006....1M}
}

@ARTICLE{Bowen2020SCamJGRA..12527813B,
       author = {{Bowen}, T.~A. and {Bale}, S.~D. and {Bonnell}, J.~W. and {Dudok de Wit}, T. and {Goetz}, K. and {Goodrich}, K. and {Gruesbeck}, J. and {Harvey}, P.~R. and {Jannet}, G. and {Koval}, A. and {MacDowall}, R.~J. and {Malaspina}, D.~M. and {Pulupa}, M. and {Revillet}, C. and {Sheppard}, D. and {Szabo}, A.},
        title = "{A Merged Search-Coil and Fluxgate Magnetometer Data Product for Parker Solar Probe FIELDS}",
      journal = {Journal of Geophysical Research (Space Physics)},
         year = 2020,
        month = may,
       volume = {125},
       number = {5},
          eid = {e27813},
        pages = {e27813},
          doi = {10.1029/2020JA027813},
archivePrefix = {arXiv},
       eprint = {2001.04587},
 primaryClass = {astro-ph.IM},
       adsurl = {https://ui.adsabs.harvard.edu/abs/2020JGRA..12527813B}
}

@ARTICLE{SrijanMicheal2025ApJ...982...96B,
       author = {{Bharati Das}, Srijan and {Terres}, Michael},
        title = "{Recovering Ion Distribution Functions. I. Slepian Reconstruction of Velocity Distribution Functions from MMS and Solar Orbiter}",
      journal = {\apj},
         year = 2025,
        month = apr,
       volume = {982},
       number = {2},
          eid = {96},
        pages = {96},
          doi = {10.3847/1538-4357/adb6a0},
archivePrefix = {arXiv},
       eprint = {2501.17294},
 primaryClass = {astro-ph.SR},
       adsurl = {https://ui.adsabs.harvard.edu/abs/2025ApJ...982...96B}
}

@ARTICLE{Shecko2008PPCF...50l4024S,
       author = {{Schekochihin}, A.~A. and {Cowley}, S.~C. and {Dorland}, W. and {Hammett}, G.~W. and {Howes}, G.~G. and {Plunk}, G.~G. and {Quataert}, E. and {Tatsuno}, T.},
        title = "{Gyrokinetic turbulence: a nonlinear route to dissipation through phase space}",
      journal = {Plasma Physics and Controlled Fusion},
         year = 2008,
        month = dec,
       volume = {50},
       number = {12},
          eid = {124024},
        pages = {124024},
          doi = {10.1088/0741-3335/50/12/124024},
archivePrefix = {arXiv},
       eprint = {0806.1069},
 primaryClass = {physics.plasm-ph},
       adsurl = {https://ui.adsabs.harvard.edu/abs/2008PPCF...50l4024S}
}

@ARTICLE{KP2009JGRA..114.0D04K,
       author = {{Kaufmann}, Richard L. and {Paterson}, William R.},
        title = "{Boltzmann H function and entropy in the plasma sheet}",
      journal = {Journal of Geophysical Research (Space Physics)},
         year = 2009,
        month = sep,
       volume = {114},
       number = {A9},
          eid = {A00D04},
        pages = {A00D04},
          doi = {10.1029/2008JA014030},
       adsurl = {https://ui.adsabs.harvard.edu/abs/2009JGRA..114.0D04K}
}

@ARTICLE{Sioulas2022ApJ...934..143S,
       author = {{Sioulas}, Nikos and {Huang}, Zesen and {Velli}, Marco and {Chhiber}, Rohit and {Cuesta}, Manuel E. and {Shi}, Chen and {Matthaeus}, William H. and {Bandyopadhyay}, Riddhi and {Vlahos}, Loukas and {Bowen}, Trevor A. and {Qudsi}, Ramiz A. and {Bale}, Stuart D. and {Owen}, Christopher J. and {Louarn}, P. and {Fedorov}, A. and {Maksimovi{\'c}}, Milan and {Stevens}, Michael L. and {Case}, Anthony and {Kasper}, Justin and {Larson}, Davin and {Pulupa}, Marc and {Livi}, Roberto},
        title = "{Magnetic Field Intermittency in the Solar Wind: Parker Solar Probe and SolO Observations Ranging from the Alfv{\'e}n Region up to 1 AU}",
      journal = {\apj},
         year = 2022,
        month = aug,
       volume = {934},
       number = {2},
          eid = {143},
        pages = {143},
          doi = {10.3847/1538-4357/ac7aa2},
archivePrefix = {arXiv},
       eprint = {2206.00871},
 primaryClass = {astro-ph.SR},
       adsurl = {https://ui.adsabs.harvard.edu/abs/2022ApJ...934..143S}
}

@ARTICLE{podesta2009ApJ...698..986P,
       author = {{Podesta}, J.~J.},
        title = "{Dependence of Solar-Wind Power Spectra on the Direction of the Local Mean Magnetic Field}",
      journal = {\apj},
         year = 2009,
        month = jun,
       volume = {698},
       number = {2},
        pages = {986-999},
          doi = {10.1088/0004-637X/698/2/986},
archivePrefix = {arXiv},
       eprint = {0901.4940},
 primaryClass = {astro-ph.EP},
       adsurl = {https://ui.adsabs.harvard.edu/abs/2009ApJ...698..986P}
}

@ARTICLE{Chen2016JPlPh..82f5302C,
       author = {{Chen}, C.~H.~K.},
        title = "{Recent progress in astrophysical plasma turbulence from solar wind observations}",
      journal = {Journal of Plasma Physics},
         year = 2016,
        month = dec,
       volume = {82},
       number = {6},
          eid = {535820602},
        pages = {535820602},
          doi = {10.1017/S0022377816001124},
archivePrefix = {arXiv},
       eprint = {1611.03386},
 primaryClass = {physics.plasm-ph},
       adsurl = {https://ui.adsabs.harvard.edu/abs/2016JPlPh..82f5302C}
}

@ARTICLE{chen2010ApJ...711L..79C,
       author = {{Chen}, C.~H.~K. and {Wicks}, R.~T. and {Horbury}, T.~S. and {Schekochihin}, A.~A.},
        title = "{Interpreting Power Anisotropy Measurements in Plasma Turbulence}",
      journal = {\apjl},
         year = 2010,
        month = mar,
       volume = {711},
       number = {2},
        pages = {L79-L83},
          doi = {10.1088/2041-8205/711/2/L79},
archivePrefix = {arXiv},
       eprint = {0909.2683},
 primaryClass = {physics.plasm-ph},
       adsurl = {https://ui.adsabs.harvard.edu/abs/2010ApJ...711L..79C}
}

@ARTICLE{Sioulas2023ApJ...951..141S,
       author = {{Sioulas}, Nikos and {Velli}, Marco and {Huang}, Zesen and {Shi}, Chen and {Bowen}, Trevor A. and {Chandran}, B.~D.~G. and {Liodis}, Ioannis and {Davis}, Nooshin and {Bale}, Stuart D. and {Horbury}, T.~S. and {Dudok de Wit}, Thierry and {Larson}, Davin and {Stevens}, Michael L. and {Kasper}, Justin and {Owen}, Christopher J. and {Case}, Anthony and {Pulupa}, Marc and {Malaspina}, David M. and {Livi}, Roberto and {Goetz}, Keith and {Harvey}, Peter R. and {MacDowall}, Robert J. and {Bonnell}, John W.},
        title = "{On the Evolution of the Anisotropic Scaling of Magnetohydrodynamic Turbulence in the Inner Heliosphere}",
      journal = {\apj},
         year = 2023,
        month = jul,
       volume = {951},
       number = {2},
          eid = {141},
        pages = {141},
          doi = {10.3847/1538-4357/acc658},
archivePrefix = {arXiv},
       eprint = {2301.03896},
 primaryClass = {physics.space-ph},
       adsurl = {https://ui.adsabs.harvard.edu/abs/2023ApJ...951..141S}
}

@ARTICLE{Horbury2008PhRvL,
       author = {{Horbury}, Timothy S. and {Forman}, Miriam and {Oughton}, Sean},
        title = "{Anisotropic Scaling of Magnetohydrodynamic Turbulence}",
      journal = {\prl},
         year = 2008,
        month = oct,
       volume = {101},
       number = {17},
          eid = {175005},
        pages = {175005},
          doi = {10.1103/PhysRevLett.101.175005},
archivePrefix = {arXiv},
       eprint = {0807.3713},
 primaryClass = {physics.plasm-ph},
       adsurl = {https://ui.adsabs.harvard.edu/abs/2008PhRvL.101q5005H}
}

@ARTICLE{Horbury2012SSRv,
       author = {{Horbury}, T.~S. and {Wicks}, R.~T. and {Chen}, C.~H.~K.},
        title = "{Anisotropy in Space Plasma Turbulence: Solar Wind Observations}",
      journal = {\ssr},
         year = 2012,
        month = nov,
       volume = {172},
       number = {1-4},
        pages = {325-342},
          doi = {10.1007/s11214-011-9821-9},
       adsurl = {https://ui.adsabs.harvard.edu/abs/2012SSRv..172..325H}
}

@ARTICLE{TrottaLarosa2024ApJ,
       author = {{Trotta}, Domenico and {Larosa}, Andrea and {Nicolaou}, Georgios and {Horbury}, Timothy S. and {Matteini}, Lorenzo and {Hietala}, Heli and {Blanco-Cano}, Xochitl and {Franci}, Luca and {Chen}, C.~H.~K. and {Zhao}, Lingling and {Zank}, Gary P. and {Cohen}, Christina M.~S. and {Bale}, Stuart D. and {Laker}, Ronan and {Fargette}, Nais and {Valentini}, Francesco and {Khotyaintsev}, Yuri and {Kieokaew}, Rungployphan and {Raouafi}, Nour and {Davies}, Emma and {Vainio}, Rami and {Dresing}, Nina and {Kilpua}, Emilia and {Karlsson}, Tomas and {Owen}, Christopher J. and {Wimmer-Schweingruber}, Robert F.},
        title = "{Properties of an Interplanetary Shock Observed at 0.07 and 0.7 au by Parker Solar Probe and Solar Orbiter}",
      journal = {\apj},
         year = 2024,
        month = feb,
       volume = {962},
       number = {2},
          eid = {147},
        pages = {147},
          doi = {10.3847/1538-4357/ad187d},
archivePrefix = {arXiv},
       eprint = {2312.05983},
 primaryClass = {astro-ph.SR},
       adsurl = {https://ui.adsabs.harvard.edu/abs/2024ApJ...962..147T}
}

@ARTICLE{Bowen2020ApJS,
       author = {{Bowen}, Trevor A. and {Mallet}, Alfred and {Huang}, Jia and {Klein}, Kristopher G. and {Malaspina}, David M. and {Stevens}, Michael and {Bale}, Stuart D. and {Bonnell}, J.~W. and {Case}, Anthony W. and {Chandran}, Benjamin D.~G. and {Chaston}, C.~C. and {Chen}, Christopher H.~K. and {Dudok de Wit}, Thierry and {Goetz}, Keith and {Harvey}, Peter R. and {Howes}, Gregory G. and {Kasper}, J.~C. and {Korreck}, Kelly E. and {Larson}, Davin and {Livi}, Roberto and {MacDowall}, Robert J. and {McManus}, Michael D. and {Pulupa}, Marc and {Verniero}, J.~L. and {Whittlesey}, Phyllis},
        title = "{Ion-scale Electromagnetic Waves in the Inner Heliosphere}",
      journal = {\apjs},
         year = 2020,
        month = feb,
       volume = {246},
       number = {2},
          eid = {66},
        pages = {66},
          doi = {10.3847/1538-4365/ab6c65},
archivePrefix = {arXiv},
       eprint = {1912.02361},
 primaryClass = {astro-ph.SR},
       adsurl = {https://ui.adsabs.harvard.edu/abs/2020ApJS..246...66B}
}

@ARTICLE{Celebre2023PhPl,
       author = {{Celebre}, G. and {Servidio}, S. and {Valentini}, F.},
        title = "{Phase space dynamics of unmagnetized plasmas: Collisionless and collisional regimes}",
      journal = {Physics of Plasmas},
         year = 2023,
        month = sep,
       volume = {30},
       number = {9},
          eid = {092304},
        pages = {092304},
          doi = {10.1063/5.0160549},
archivePrefix = {arXiv},
       eprint = {2306.03567},
 primaryClass = {physics.plasm-ph},
       adsurl = {https://ui.adsabs.harvard.edu/abs/2023PhPl...30i2304C}
}

@ARTICLE{Cerri2021ApJ,
       author = {{Cerri}, S.~S. and {Arzamasskiy}, L. and {Kunz}, M.~W.},
        title = "{On Stochastic Heating and Its Phase-space Signatures in Low-beta Kinetic Turbulence}",
      journal = {\apj},
         year = 2021,
        month = aug,
       volume = {916},
       number = {2},
          eid = {120},
        pages = {120},
          doi = {10.3847/1538-4357/abfbde},
archivePrefix = {arXiv},
       eprint = {2102.09654},
 primaryClass = {astro-ph.SR},
       adsurl = {https://ui.adsabs.harvard.edu/abs/2021ApJ...916..120C}
}

@ARTICLE{Pezzi2019ApJ,
       author = {{Pezzi}, O. and {Perrone}, D. and {Servidio}, S. and {Valentini}, F. and {Sorriso-Valvo}, L. and {Veltri}, P.},
        title = "{Proton-Proton Collisions in the Turbulent Solar Wind: Hybrid Boltzmann-Maxwell Simulations}",
      journal = {\apj},
         year = 2019,
        month = dec,
       volume = {887},
       number = {2},
          eid = {208},
        pages = {208},
          doi = {10.3847/1538-4357/ab5285},
archivePrefix = {arXiv},
       eprint = {1903.03398},
 primaryClass = {physics.plasm-ph},
       adsurl = {https://ui.adsabs.harvard.edu/abs/2019ApJ...887..208P}
}

@ARTICLE{kp2009JGRA,
       author = {{Kaufmann}, Richard L. and {Paterson}, William R.},
        title = "{Boltzmann H function and entropy in the plasma sheet}",
      journal = {Journal of Geophysical Research (Space Physics)},
         year = 2009,
        month = sep,
       volume = {114},
       number = {A9},
          eid = {A00D04},
        pages = {A00D04},
          doi = {10.1029/2008JA014030},
       adsurl = {https://ui.adsabs.harvard.edu/abs/2009JGRA..114.0D04K}
}

@ARTICLE{Pezzi2021MNRAS,
       author = {{Pezzi}, O. and {Liang}, H. and {Juno}, J.~L. and {Cassak}, P.~A. and {V{\'a}sconez}, C.~L. and {Sorriso-Valvo}, L. and {Perrone}, D. and {Servidio}, S. and {Roytershteyn}, V. and {TenBarge}, J.~M. and {Matthaeus}, W.~H.},
        title = "{Dissipation measures in weakly collisional plasmas}",
      journal = {\mnras},
         year = 2021,
        month = aug,
       volume = {505},
       number = {4},
        pages = {4857-4873},
          doi = {10.1093/mnras/stab1516},
archivePrefix = {arXiv},
       eprint = {2101.00722},
 primaryClass = {physics.plasm-ph},
       adsurl = {https://ui.adsabs.harvard.edu/abs/2021MNRAS.505.4857P}
}

@ARTICLE{Coburn2024ApJ,
       author = {{Coburn}, Jesse T. and {Verscharen}, Daniel and {Owen}, Christopher J. and {Maksimovic}, Milan and {Horbury}, Timothy S. and {Chen}, Christopher H.~K. and {Guo}, Fan and {Fu}, Xiangrong and {Liu}, Jingting and {Abraham}, Joel B. and {Nicolaou}, Georgios and {Innocenti}, Maria Elena and {Micera}, Alfredo and {Jagarlamudi}, Vamsee Krishna},
        title = "{The Regulation of the Solar Wind Electron Heat Flux by Wave{\textendash}Particle Interactions}",
      journal = {\apj},
         year = 2024,
        month = mar,
       volume = {964},
       number = {1},
          eid = {100},
        pages = {100},
          doi = {10.3847/1538-4357/ad1329},
       adsurl = {https://ui.adsabs.harvard.edu/abs/2024ApJ...964..100C}
}

@ARTICLE{Cerri2018ApJ,
       author = {{Cerri}, S.~S. and {Kunz}, M.~W. and {Califano}, F.},
        title = "{Dual Phase-space Cascades in 3D Hybrid-Vlasov-Maxwell Turbulence}",
      journal = {\apjl},
         year = 2018,
        month = mar,
       volume = {856},
       number = {1},
          eid = {L13},
        pages = {L13},
          doi = {10.3847/2041-8213/aab557},
archivePrefix = {arXiv},
       eprint = {1802.06133},
 primaryClass = {physics.plasm-ph},
       adsurl = {https://ui.adsabs.harvard.edu/abs/2018ApJ...856L..13C}
}

@ARTICLE{Verniero2022ApJ,
       author = {{Verniero}, J.~L. and {Chandran}, B.~D.~G. and {Larson}, D.~E. and {Paulson}, K. and {Alterman}, B.~L. and {Badman}, S. and {Bale}, S.~D. and {Bonnell}, J.~W. and {Bowen}, T.~A. and {de Wit}, T. Dudok and {Kasper}, J.~C. and {Klein}, K.~G. and {Lichko}, E. and {Livi}, R. and {McManus}, M.~D. and {Rahmati}, A. and {Verscharen}, D. and {Walters}, J. and {Whittlesey}, P.~L.},
        title = "{Strong Perpendicular Velocity-space Diffusion in Proton Beams Observed by Parker Solar Probe}",
      journal = {\apj},
         year = 2022,
        month = jan,
       volume = {924},
       number = {2},
          eid = {112},
        pages = {112},
          doi = {10.3847/1538-4357/ac36d5},
archivePrefix = {arXiv},
       eprint = {2110.08912},
 primaryClass = {astro-ph.SR},
       adsurl = {https://ui.adsabs.harvard.edu/abs/2022ApJ...924..112V}
}

@ARTICLE{Verniero2020ApJS,
       author = {{Verniero}, J.~L. and {Larson}, D.~E. and {Livi}, R. and {Rahmati}, A. and {McManus}, M.~D. and {Pyakurel}, P. Sharma and {Klein}, K.~G. and {Bowen}, T.~A. and {Bonnell}, J.~W. and {Alterman}, B.~L. and {Whittlesey}, P.~L. and {Malaspina}, David M. and {Bale}, S.~D. and {Kasper}, J.~C. and {Case}, A.~W. and {Goetz}, K. and {Harvey}, P.~R. and {Korreck}, K.~E. and {MacDowall}, R.~J. and {Pulupa}, M. and {Stevens}, M.~L. and {de Wit}, T. Dudok},
        title = "{Parker Solar Probe Observations of Proton Beams Simultaneous with Ion-scale Waves}",
      journal = {\apjs},
         year = 2020,
        month = may,
       volume = {248},
       number = {1},
          eid = {5},
        pages = {5},
          doi = {10.3847/1538-4365/ab86af},
archivePrefix = {arXiv},
       eprint = {2004.03009},
 primaryClass = {physics.space-ph},
       adsurl = {https://ui.adsabs.harvard.edu/abs/2020ApJS..248....5V}
}

@ARTICLE{Bowen2022PhRvL,
       author = {{Bowen}, Trevor A. and {Chandran}, Benjamin D.~G. and {Squire}, Jonathan and {Bale}, Stuart D. and {Duan}, Die and {Klein}, Kristopher G. and {Larson}, Davin and {Mallet}, Alfred and {McManus}, Michael D. and {Meyrand}, Romain and {Verniero}, Jaye L. and {Woodham}, Lloyd D.},
        title = "{In Situ Signature of Cyclotron Resonant Heating in the Solar Wind}",
      journal = {\prl},
         year = 2022,
        month = oct,
       volume = {129},
       number = {16},
          eid = {165101},
        pages = {165101},
          doi = {10.1103/PhysRevLett.129.165101},
archivePrefix = {arXiv},
       eprint = {2111.05400},
 primaryClass = {astro-ph.SR},
       adsurl = {https://ui.adsabs.harvard.edu/abs/2022PhRvL.129p5101B}
}

@ARTICLE{FoxVelli2016,
       author = {{Fox}, N.~J. and {Velli}, M.~C. and {Bale}, S.~D. and {Decker}, R. and {Driesman}, A. and {Howard}, R.~A. and {Kasper}, J.~C. and {Kinnison}, J. and {Kusterer}, M. and {Lario}, D. and {Lockwood}, M.~K. and {McComas}, D.~J. and {Raouafi}, N.~E. and {Szabo}, A.},
        title = "{The Solar Probe Plus Mission: Humanity's First Visit to Our Star}",
      journal = {\ssr},
         year = 2016,
        month = dec,
       volume = {204},
       number = {1-4},
        pages = {7-48},
          doi = {10.1007/s11214-015-0211-6},
       adsurl = {https://ui.adsabs.harvard.edu/abs/2016SSRv..204....7F}
}

@ARTICLE{Livi2022ApJ,
       author = {{Livi}, Roberto and {Larson}, Davin E. and {Kasper}, Justin C. and {Abiad}, Robert and {Case}, A.~W. and {Klein}, Kristopher G. and {Curtis}, David W. and {Dalton}, Gregory and {Stevens}, Michael and {Korreck}, Kelly E. and {Ho}, George and {Robinson}, Miles and {Tiu}, Chris and {Whittlesey}, Phyllis L. and {Verniero}, Jaye L. and {Halekas}, Jasper and {McFadden}, James and {Marckwordt}, Mario and {Slagle}, Amanda and {Abatcha}, Mamuda and {Rahmati}, Ali and {McManus}, Michael D.},
        title = "{The Solar Probe ANalyzer-Ions on the Parker Solar Probe}",
      journal = {\apj},
         year = 2022,
        month = oct,
       volume = {938},
       number = {2},
          eid = {138},
        pages = {138},
          doi = {10.3847/1538-4357/ac93f5},
       adsurl = {https://ui.adsabs.harvard.edu/abs/2022ApJ...938..138L}
}

@ARTICLE{phyllisSpanE2020,
       author = {{Whittlesey}, Phyllis L. and {Larson}, Davin E. and {Kasper}, Justin C. and {Halekas}, Jasper and {Abatcha}, Mamuda and {Abiad}, Robert and {Berthomier}, M. and {Case}, A.~W. and {Chen}, Jianxin and {Curtis}, David W. and {Dalton}, Gregory and {Klein}, Kristopher G. and {Korreck}, Kelly E. and {Livi}, Roberto and {Ludlam}, Michael and {Marckwordt}, Mario and {Rahmati}, Ali and {Robinson}, Miles and {Slagle}, Amanda and {Stevens}, M.~L. and {Tiu}, Chris and {Verniero}, J.~L.},
        title = "{The Solar Probe ANalyzers{\textemdash}Electrons on the Parker Solar Probe}",
      journal = {\apjs},
         year = 2020,
        month = feb,
       volume = {246},
       number = {2},
          eid = {74},
        pages = {74},
          doi = {10.3847/1538-4365/ab7370},
archivePrefix = {arXiv},
       eprint = {2002.04080},
 primaryClass = {astro-ph.IM},
       adsurl = {https://ui.adsabs.harvard.edu/abs/2020ApJS..246...74W}
}

@ARTICLE{Kasper2016,
       author = {{Kasper}, Justin C. and {Abiad}, Robert and {Austin}, Gerry and {Balat-Pichelin}, Marianne and {Bale}, Stuart D. and {Belcher}, John W. and {Berg}, Peter and {Bergner}, Henry and {Berthomier}, Matthieu and {Bookbinder}, Jay and {Brodu}, Etienne and {Caldwell}, David and {Case}, Anthony W. and {Chandran}, Benjamin D.~G. and {Cheimets}, Peter and {Cirtain}, Jonathan W. and {Cranmer}, Steven R. and {Curtis}, David W. and {Daigneau}, Peter and {Dalton}, Greg and {Dasgupta}, Brahmananda and {DeTomaso}, David and {Diaz-Aguado}, Millan and {Djordjevic}, Blagoje and {Donaskowski}, Bill and {Effinger}, Michael and {Florinski}, Vladimir and {Fox}, Nichola and {Freeman}, Mark and {Gallagher}, Dennis and {Gary}, S. Peter and {Gauron}, Tom and {Gates}, Richard and {Goldstein}, Melvin and {Golub}, Leon and {Gordon}, Dorothy A. and {Gurnee}, Reid and {Guth}, Giora and {Halekas}, Jasper and {Hatch}, Ken and {Heerikuisen}, Jacob and {Ho}, George and {Hu}, Qiang and {Johnson}, Greg and {Jordan}, Steven P. and {Korreck}, Kelly E. and {Larson}, Davin and {Lazarus}, Alan J. and {Li}, Gang and {Livi}, Roberto and {Ludlam}, Michael and {Maksimovic}, Milan and {McFadden}, James P. and {Marchant}, William and {Maruca}, Bennet A. and {McComas}, David J. and {Messina}, Luciana and {Mercer}, Tony and {Park}, Sang and {Peddie}, Andrew M. and {Pogorelov}, Nikolai and {Reinhart}, Matthew J. and {Richardson}, John D. and {Robinson}, Miles and {Rosen}, Irene and {Skoug}, Ruth M. and {Slagle}, Amanda and {Steinberg}, John T. and {Stevens}, Michael L. and {Szabo}, Adam and {Taylor}, Ellen R. and {Tiu}, Chris and {Turin}, Paul and {Velli}, Marco and {Webb}, Gary and {Whittlesey}, Phyllis and {Wright}, Ken and {Wu}, S.~T. and {Zank}, Gary},
        title = "{Solar Wind Electrons Alphas and Protons (SWEAP) Investigation: Design of the Solar Wind and Coronal Plasma Instrument Suite for Solar Probe Plus}",
      journal = {\ssr},
         year = 2016,
        month = dec,
       volume = {204},
       number = {1-4},
        pages = {131-186},
          doi = {10.1007/s11214-015-0206-3},
       adsurl = {https://ui.adsabs.harvard.edu/abs/2016SSRv..204..131K}
}
\bibliographystyle{aasjournal}
\end{document}